\documentclass[11pt,a4paper]{article} 
\usepackage{jheppub}

\usepackage{latexsym}
\usepackage{amssymb}

\numberwithin{equation}{section}
\allowdisplaybreaks

\newcommand{\as}{\alpha_s}

\newcommand{\be}{\begin{equation}}
\newcommand{\ee}{\end{equation}}

\newcommand{\bea}{\begin{eqnarray}}
\newcommand{\eea}{\end{eqnarray}}
\newcommand{\bean}{\begin{eqnarray*}}
\newcommand{\eean}{\end{eqnarray*}}

\newcommand{\half} {\frac{1}{2}}
\newcommand{\om} {\omega}

\newcommand{\pv}{{\mathbf p}}
\newcommand{\kv}{{\mathbf k}}
\newcommand{\nv}{{\mathbf n}}

\newcommand{\vecnul}{{\mathbf 0}}

\title{
$S$ wave bottomonium states moving in a quark-gluon plasma from lattice NRQCD
}

\author[a]{G.~Aarts,}
\author[a]{ C.~Allton,}
\author[b]{S.~Kim,} 
\author[c]{M.~P.~Lombardo,}
\author[d]{M.~B.~Oktay,}
\author[e]{S.~M.~Ryan,}
\author[f]{D.~K.~Sinclair,}
\author[g]{and  J.-I.~Skullerud}

\affiliation[a]{Department of Physics, Swansea University, Swansea, United Kingdom} 
\affiliation[b]{Department of Physics, Sejong University, Seoul 143-747, Korea} 
\affiliation[c]{INFN-Laboratori Nazionali di Frascati, I-00044, Frascati (RM) Italy} 
\affiliation[d]{Physics Department, University of Utah, Salt Lake City,  Utah, USA} 
\affiliation[e]{School of Mathematics, Trinity College, Dublin 2, Ireland} 
\affiliation[f]{HEP Division, Argonne National Laboratory, 9700 South Cass Avenue, Argonne, Illinois 60439, USA} 
\affiliation[g]{Department of Mathematical Physics, National University of Ireland Maynooth, Maynooth, County Kildare, Ireland}

\emailAdd{g.aarts@swan.ac.uk}
\emailAdd{c.allton@swansea.ac.uk}
\emailAdd{skim@sejong.ac.kr}
\emailAdd{Mariapaola.Lombardo@lnf.infn.it}
\emailAdd{oktay@physics.utah.edu}
\emailAdd{ryan@maths.tcd.ie}
\emailAdd{dks@hep.anl.gov}
\emailAdd{jonivar@thphys.nuim.ie}

\abstract{
 We extend our study of bottomonium spectral functions in the quark-gluon plasma to nonzero momentum. We use lattice QCD simulations with two flavours of light quark on highly anisotropic lattices and treat the bottom quark with nonrelativistic QCD (NRQCD). 
We focus on $S$ wave ($\Upsilon$ and $\eta_b$) channels  and consider nonrelativistic velocities, $v/c\lesssim 0.2$.
A comparison with predictions from effective field theory is made.
 }

\keywords{Lattice QCD, Thermal Field Theory}

\arxivnumber{1210.2903}

\begin{document}
\maketitle

\section{Introduction}
 \label{sec:intro}
 
 Quarkonia, heavy quark--antiquark bound states, play an important role as probes of the quark-gluon plasma (QGP) created in relativistic heavy-ion collisions at RHIC and the LHC \cite{Matsui:1986dk} (for reviews, see e.g.\ refs.\  \cite{Satz:2006kba,Rapp:2008tf}).
 How the width of a particular bound state -- and its dissociation rate -- varies with temperature,
depends on the quark content (charmonium,
bottomomium) and on its quantum numbers (e.g.\ $S$ or $P$ wave, in
nonrelativistic notation). Recently the CMS collaboration has observed
sequential Upsilon suppression in PbPb collisions at the LHC
\cite{Chatrchyan:2011pe,:2012fr}, which has inspired significant
phenomenological and theoretical activity 
\cite{Strickland:2011mw,Strickland:2011aa,Brezinski:2011ju,Emerick:2011xu,Song:2011ev,Dutta:2012nw}.

Traditionally, quarkonium suppression has been studied using potential models (see refs.\ \cite{Mocsy:2007yj,Mocsy:2007jz} and references therein) and with lattice QCD computations of quarkonium spectral functions \cite{Umeda:2002vr,Asakawa:2003re,Datta:2003ww,Jakovac:2006sf,Aarts:2007pk,Ding:2012sp,Ohno:2011zc} (recent reviews can be found in refs.\  \cite{Ding:2012ar,Kaczmarek:2012ne}).
Several years ago the theoretical study of quarkonia in a thermal medium was formulated in a more systematic fashion by casting the problem in the language of effective field theory (EFT)
\cite{Laine:2006ns,Laine:2007gj,Burnier:2007qm,Brambilla:2008cx,Brambilla:2010vq,Brambilla:2011sg,Beraudo:2007ky,Beraudo:2010tw,Escobedo:2011ie}. Besides the energy scales  available in vacuum: the heavy quark mass $M_q$, the inverse system size $1/a_0\sim M_q\as$, and the binding energy $E_b\sim M_q\as^2$ (with $\as$ the strong coupling constant), new thermal scales are provided by the temperature $T$ and the inverse Debye screening length $1/r_D\sim \sqrt\as T$. The relevance of these scales  depends on the temperature of the QGP and the magnitude of the strong coupling constant.
In analytical studies,  $\as$ is usually assumed to be small enough for a hierarchy of scales and EFTs to emerge.
An important conceptual finding was the emergence of a complex potential at nonzero temperature
\cite{Laine:2006ns}, which has stimulated further complex potential model studies
\cite{Petreczky:2010tk,Margotta:2011ta} as well as attempts to extract the complex potential from lattice QCD  \cite{Rothkopf:2011db,Burnier:2012ib}.

For temperatures reached in heavy-ion collisions and in the case of bottomonium, the heavy quark mass is the highest energy scale present. Integrating out this scale yields nonrelativistic QCD (NRQCD) \cite{Caswell:1985ui,Lepage:1992tx,Bodwin:1994jh,Brambilla:2004jw}, just as in vacuum.
Recently we have employed lattice NRQCD
\cite{Davies:1994mp,Davies:1995db,Davies:1998im} to study the fate of
$P$ wave \cite{Aarts:2010ek} and $S$ wave
\cite{Aarts:2010ek,Aarts:2011sm} bottomonium states nonperturbatively,
using lattice QCD simulations of nonrelativistic bottom quarks,
propagating through a medium of thermal gluons and  two light
flavours, at temperatures between $0.4T_c$ and $2.1T_c$. We found that the
use of NRQCD greatly enhances the signature for quarkonium
melting/survival, since it avoids 
 several problems which have complicated the study of
  relativistic quarks in thermal equilibrium
 \cite{Aarts:2002cc,Umeda:2007hy,Aarts:2007wj,Petreczky:2008px}.

Our studies indicate that bound states in the $P$ wave channels ($\chi_b, h_b$) melt quickly as the temperature is raised above $T_c$, while the signal in the $S$ wave channels ($\Upsilon, \eta_b$) is consistent with survival \cite{Aarts:2010ek}. A closer study of the $S$ wave spectral functions, constructed using the maximum entropy method (MEM) \cite{Asakawa:2000tr,Bryan}, subsequently showed that the ground state peaks appear  to survive, whereas excited states are suppressed, consistent with the recent CMS results
\cite{Chatrchyan:2011pe,:2012fr}.
Moreover, medium effects in the ground state peak can be captured by a temperature-dependent position and width \cite{Aarts:2011sm}. The thermal mass shift and width were found to be consistent with those predicted by EFT calculations \cite{Brambilla:2010vq}, assuming $\as\sim 0.4$ \cite{Aarts:2011sm}.
 
 In this paper we extend the lattice calculation and consider bottomonium $S$ wave correlators at nonzero momentum.
 While we find a significant momentum dependence in the euclidean correlators (as expected), we observe that this momentum dependence is to a large extent independent of the temperature. This result is further investigated by constructing the corresponding spectral functions at nonzero momentum and extracting the momentum-dependent position and width of the ground state peak. We note here that  in the literature only a handful of lattice studies of spectral functions at nonzero momentum can be found, all using the relativistic formulation:  see ref.\ 
 \cite{Aarts:2006wt}  for light (staggered) quarks and refs.\ \cite{Oktay:2010tf,Nonaka:2010zz} for studies of charmonium.
 A  lattice study of the (gauge fixed) quark propagator at nonzero momentum can be found in ref.\ \cite{Karsch:2009tp}.
 
This paper is organised as follows. In the next section,  we first discuss how the different scales appear in our lattice simulation and, most importantly, what  ground state velocities can be reached.
The main results for the correlators and spectral functions are given in section \ref{sec:low} for the hadronic phase and in section \ref{sec:high} for the quark-gluon plasma phase, where we also comment on possible comparisons with the predictions from thermal EFT calculations at nonzero velocity. A summary is given in section \ref{sec:sum}. 
 In appendix \ref{sec:free} we consider free nonrelativistic quarks at nonzero momentum, while in appendix \ref{sec:unc} further details regarding the maximum entropy method are given.

 \section{Scales on the lattice}
 \label{sec:scales} 
 
 In this section, some simple estimates of the scales appearing in the lattice simulation and in the EFT formulation are given. 
 We consider bottomonium: for the sake of simplicity let us assume a
 quark mass $M_q\sim 5$ GeV and an $S$ wave ground state (at rest) of
 $M_S\sim 9.5$ GeV. The temperatures that we are able to reach in our
 current two-flavour, highly anisotropic lattice simulations go up to
 458 MeV. Some details of the lattice setup are given in
 table~\ref{tab:lattice}. 
   We use dynamical light Wilson-type quark flavours, with $m_\pi/m_\rho \simeq 0.54$.
    The temperature is very precisely
   determined in MeV; in units of $T_c$ less so,
     primarily because we only have a rough estimate of $T_c$ on these
     lattices.
More details can be found in refs.\ \cite{Morrin:2006tf,Aarts:2007pk,Oktay:2010tf,Aarts:2011sm}. At these temperatures, one may use $0.3\lesssim \as(T)\lesssim 0.4$ (previously we found $\as\sim 0.4$  \cite{Aarts:2011sm}, based on a comparison with EFT predictions  \cite{Brambilla:2010vq}). We arrive therefore at the hierarchy
 \be
 M_q \gg M_q\as \gg T  \sim  M_q\as^2,
 \ee
 where the final comparison depends on the size of $\as$ and on the temperature.
The use of NRQCD is therefore very well motivated.  Details of the ${\cal O}(v^4)$ improved lattice NRQCD formulation we use can be found in ref.\ \cite{Aarts:2011sm}.

\begin{table}[t]
\begin{center}
\vspace*{0.2cm}
\begin{tabular}{| l | rrrrrrr | }
\hline
$N_\tau$ 		& 80 & 32 & 28 & 24 & 20 & 18 & 16 \\
$T$(MeV) 	& 90 & 230 & 263 & 306 & 368 & 408 & 458 \\
$T/T_c$ 		& 0.42 & 1.05 & 1.20 & 1.40 & 1.68 & 1.86 & 2.09 \\ 
$N_{\rm cfg}$   & 250 & 1000 & 1000 & 500 & 1000 & 1000 & 1000 \\
\hline
\end{tabular}
\vspace*{0.2cm}
\caption{Two-flavour lattice details: the lattice size is $N_s^3\times
  N_\tau$ with $N_s=12$, and the lattice spacing is  $a_s\simeq 0.162$
  fm, $a_\tau^{-1} =7.35(3)$ GeV determined from the $1P-1S$ splitting 
in charmonium. The anisotropy is $a_s/a_\tau=6$ \cite{Aarts:2011sm}.  The temperatures in MeV
    have uncertainties of $\sim0.5\%$ from the uncertainty in the
    lattice spacing, while the temperatures in units of $T_c$ are
    rough estimates with an accuracy of $\sim10\%$.
}
\label{tab:lattice}
\end{center}
\end{table}

 The momenta and velocities that are accessible on the lattice are
 constrained by the discretization and the spatial lattice
 spacing. The lattice dispersion relation  reads
   \be
 a_s^2\pv^2 = 4\sum_{i=1}^3 \sin^2\frac{p_i}{2}, \quad\quad
   p_i = \frac{2\pi n_i}{N_s}, \quad\quad
    -\frac{N_s}{2} < n_i \leq \frac{N_s}{2}.
 \ee
   To avoid lattice artefacts, only momenta with $n_i<N_s/4$ are used: we consider the combinations (and permutations thereof) given in table \ref{tab:mom}.
   The largest momentum, using $\nv = (2,2,0)$, is $|\pv| \simeq 1.73$ GeV, corresponding to $v = |\pv|/M_S \simeq 0.2$. Therefore, the range of velocities we consider is nonrelativistic.

\begin{table}[t]
\begin{center}
\begin{tabular}{| l | ccccccc | }
\hline
$\nv$ 			& (1,0,0)	& (1,1,0)   & (1,1,1)  &  (2,0,0)	& (2,1,0)	& (2,1,1)	& (2,2,0) 	\\
$|\pv|$ (GeV)		& 0.634    & 0.900 	& 1.10 	& 1.23	& 1.38 	& 1.52 	& 1.73	\\
$v$ [$\Upsilon(^3S_1)$]	& 0.0670 	& 0.0951	& 0.116	& 0.130	& 0.146	& 0.161 	& 0.183 	\\
$v$ [$\eta_b(^1S_0)$]	& 0.0672 	& 0.0954 	& 0.117 	& 0.130 	& 0.146 	& 0.161 	& 0.183	\\
\hline
\end{tabular}
\vspace*{0.2cm}
 \caption{Nonzero momenta used in this study. Also indicated are the
   corresponding velocities $v=|\pv|/M_S$ of the ground states in the
   vector ($\Upsilon$) and pseudoscalar ($\eta_b$) channels, using the
   ground state masses determined previously  \cite{Aarts:2010ek},  $M_{\Upsilon}=9.460$ GeV and  $M_{\eta_b}=9.438$ GeV.
 }  
\label{tab:mom}
\end{center}
\end{table}

\section{Low temperature}
\label{sec:low}

\begin{figure}[h]
\begin{center}
\epsfig{figure=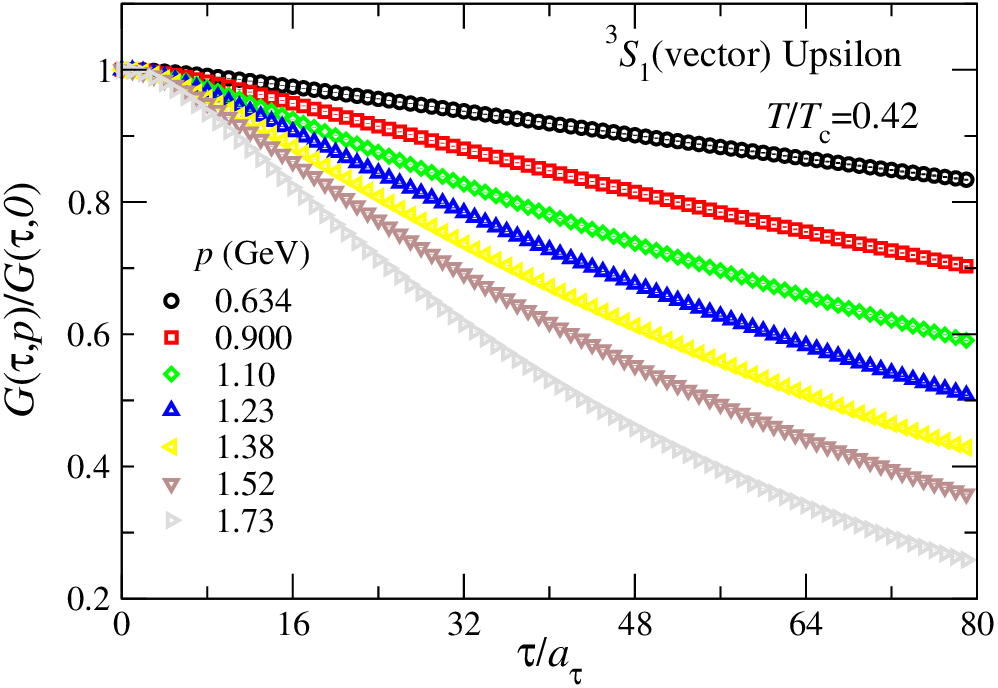,width=0.45\textwidth}
\epsfig{figure=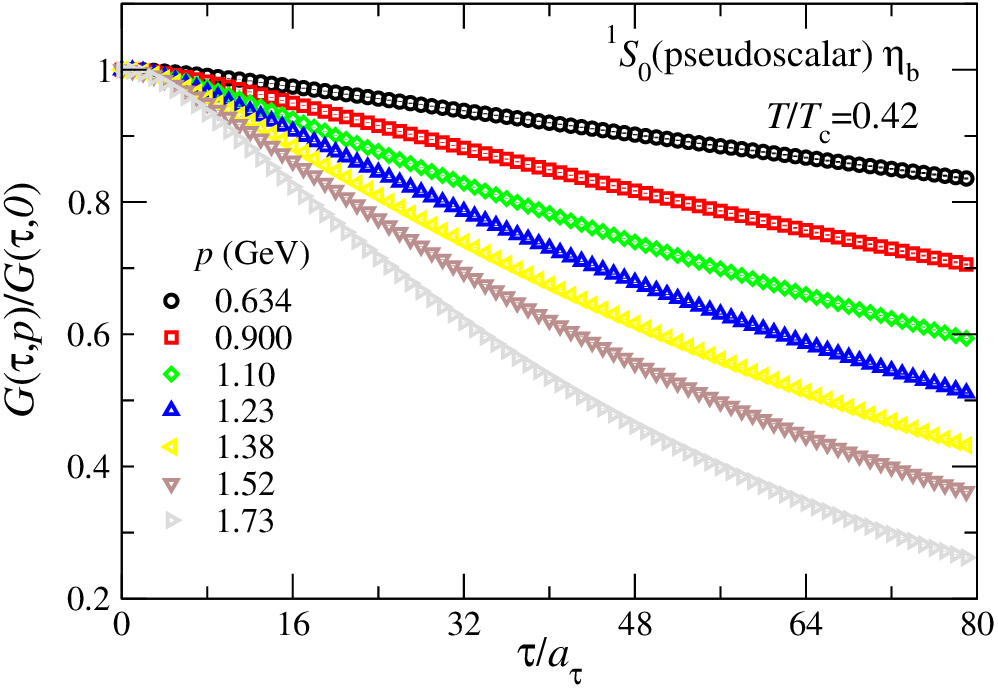,width=0.45\textwidth}
 \caption{
 Euclidean correlators at nonzero momentum $p=|\pv|$ normalized  by the  correlator at zero momentum, $G(\tau,\pv)/G(\tau,\vecnul)$, at the lowest temperature in the vector ($\Upsilon$) channel (left) and the pseudoscalar ($\eta_b$) channel (right). 
 Error bars are considerably smaller than the symbols.
}
 \label{fig:1}
\end{center}
\end{figure}

We start by presenting the results at the lowest temperature in the
hadronic phase, i.e.\ $T/T_c\sim 0.42$. Fig.\ \ref{fig:1} (left) shows
the euclidean correlation functions at nonzero momentum, normalized
by the correlator at zero momentum.  
The momentum dependence is introduced  in such a way that $G(\tau=0,\pv)=G(\tau=0,\vecnul)$ for all momenta.

We observe considerable momentum dependence. This is easily understood. At large enough euclidean times, the correlator is dominated by the ground state and 
\be
 G(\tau,\pv)\sim e^{-M(\pv)\tau}.
 \ee 
We then find
\be
\label{eq:ratio}
\frac{G(\tau,\pv)}{G(\tau,\vecnul)}  \sim  e^{-\Delta M(\pv)\tau} \approx 1-\Delta M(\pv)\tau + \ldots,
\ee
where 
\be
\Delta M(\pv) \equiv M(\pv)-M(\vecnul) \approx \frac{\pv^2}{2M_{\rm kin}} =  \half M_{\rm kin}v^2. 
\ee
Here $M_{\rm kin}$ is the kinetic mass, which is equal to the rest mass $M(\vecnul)$ when the quark mass is carefully tuned.\footnote{In our study this holds only approximately.} 
The ratios shown in fig.\ \ref{fig:1} are therefore expected to drop to zero exponentially at large 
euclidean time.
However, at a finite time and for small momenta, i.e.\ when $\Delta M(\pv)\tau\ll 1$, this translates into linear time dependence, as indicated in eq.~(\ref{eq:ratio}) and visible in fig.\ \ref{fig:1} as well.

\begin{figure}[t]
\begin{center}
\epsfig{figure=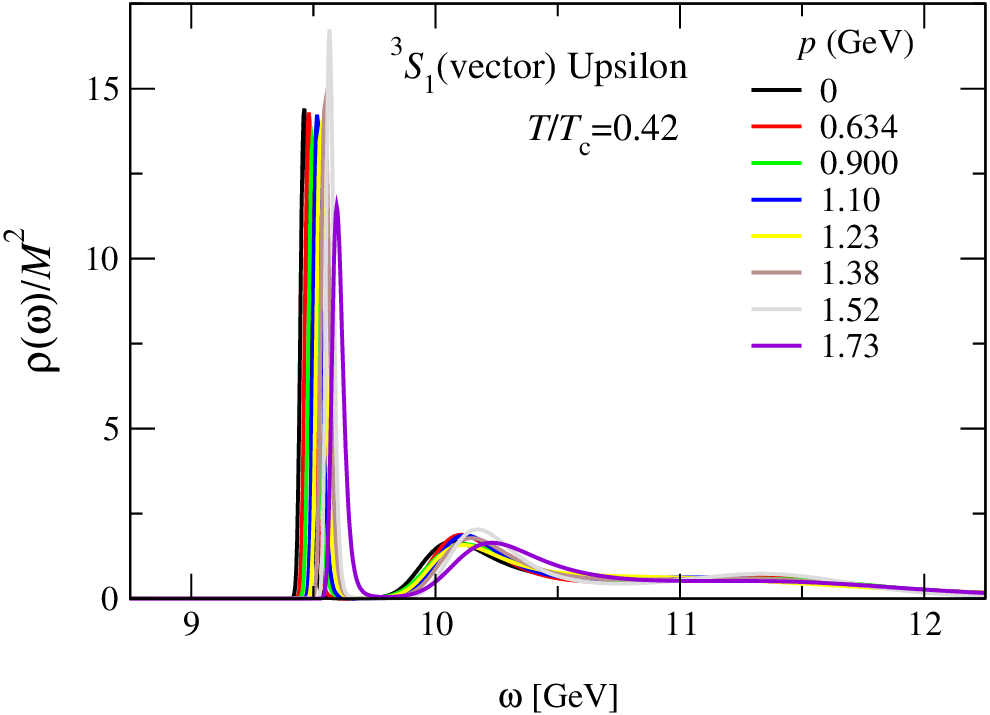,width=0.45\textwidth}
\epsfig{figure=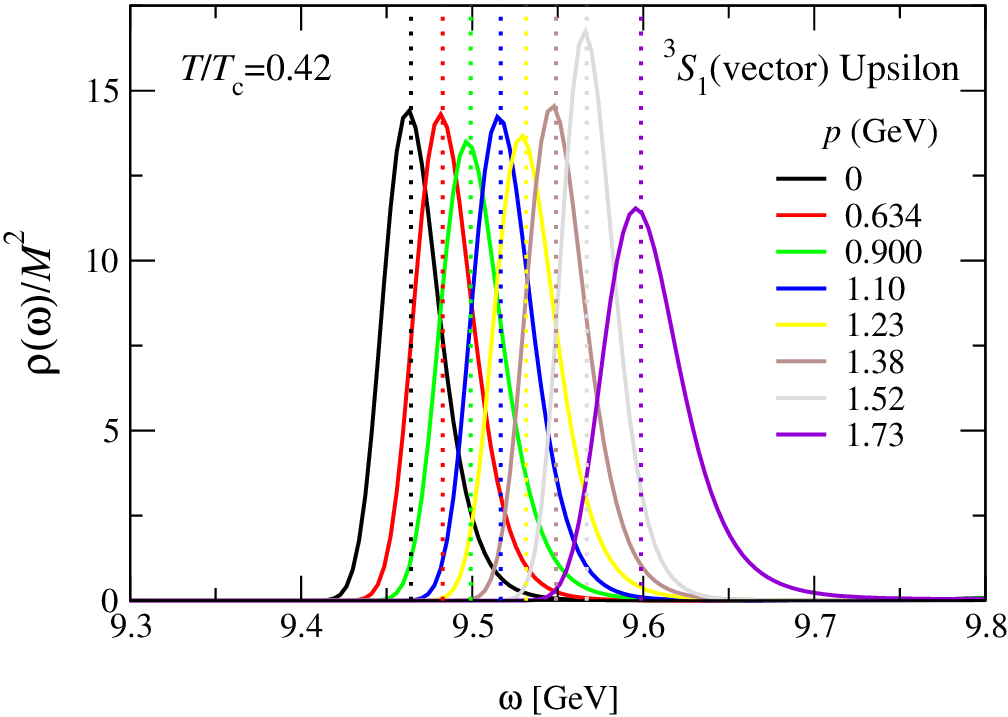,width=0.45\textwidth}
 \caption{
 Spectral functions $\rho(\om,\pv)$, normalized by the heavy quark mass, as a function of energy in the vector ($\Upsilon$) channel at the lowest temperature, for several momenta $p$.
 The graph on the right shows a close-up, with vertical lines indicating the position of the ground state at each momentum obtained via standard exponential fits.
 }
 \label{fig:rho-upsilon-80}
\end{center}
\begin{center}
\epsfig{figure=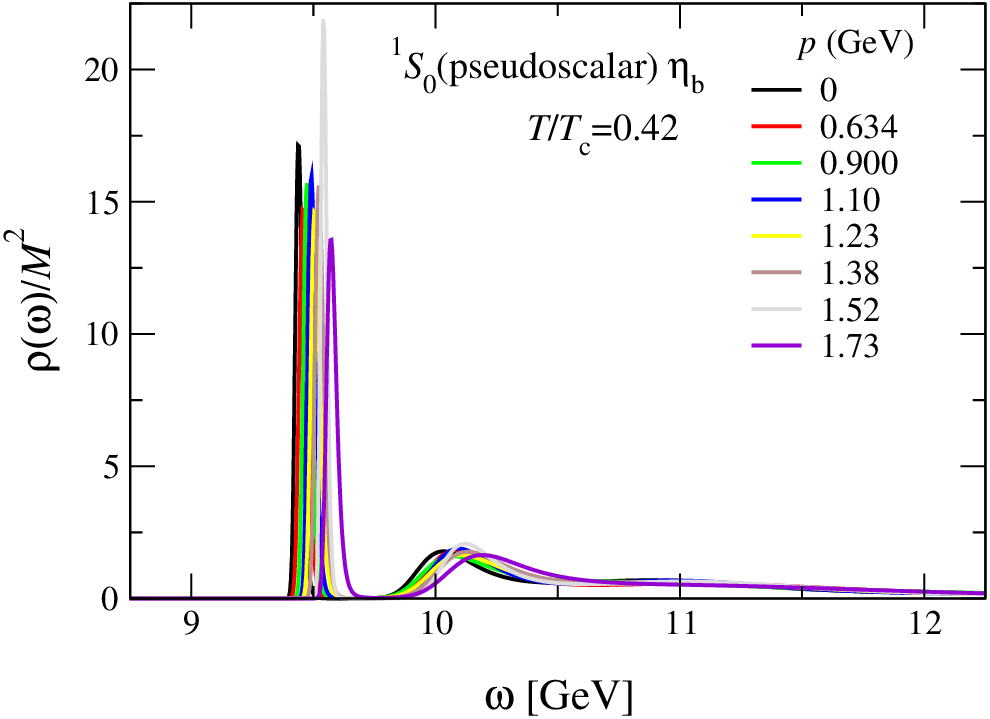,width=0.45\textwidth}
\epsfig{figure=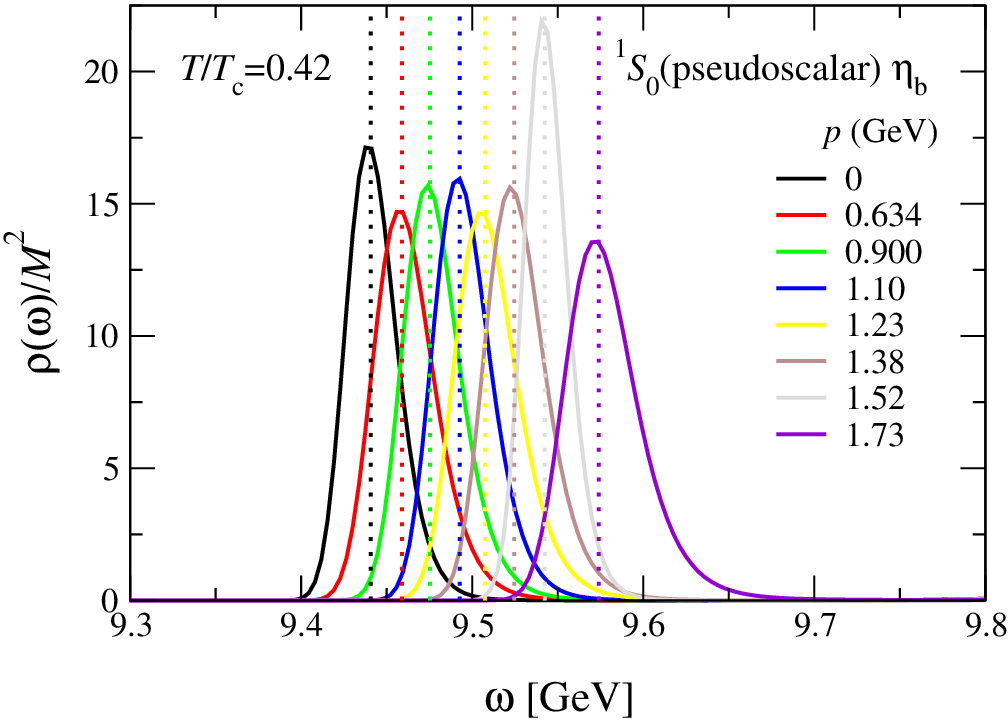,width=0.45\textwidth}
 \caption{
 As in the preceding figure, for the pseudoscalar ($\eta_b$) channel.
  }
\label{fig:rho-eta-80}
\end{center}
\end{figure}

In NRQCD, euclidean correlation functions and spectral functions  are related 
by
\be
G(\tau,\pv) = \int_{\om_{\rm min}}^{\om_{\rm max}}\frac{d\omega}{2\pi} \, K(\tau,\om)\rho(\om,\pv), \quad\quad\quad K(\tau,\om) = e^{-\om\tau},
\ee
both at zero and nonzero temperature  \cite{Burnier:2007qm,Aarts:2010ek,Aarts:2011sm}.
We construct the spectral functions using the maximum entropy method \cite{Asakawa:2000tr,Bryan}, referring to ref.\ \cite{Aarts:2011sm} for details of MEM in this context. We emphasize that in this setup the temperature dependence does not enter via the kernel $K(\tau,\om)$, but arises solely from the presence of 
the medium of gluons and light quarks at different temperatures. This greatly enhances the robustness of MEM \cite{Aarts:2011sm} and avoids 
  a number of problems associated with relativistic quarks, such
as zero-modes~\cite{Umeda:2007hy}.
Note that due to the separation of scales ($M_q\gg T$), thermal effects for the heavy quark are exponentially suppressed; 
such terms are in fact removed from the NRQCD partition function when integrating out the heavy quark energy scale.  Whether $b$ quarks actually thermalise in the quark--gluon plasma created in heavy-ion collisions at RHIC and LHC is a separate question which we will not address here. 

The spectral functions are shown in figs.\ \ref{fig:rho-upsilon-80} and \ref{fig:rho-eta-80} 
for the vector and the pseudoscalar channel. We observe that the ground state peak shifts to the right with increasing momentum, as expected. 
The vertical dashed lines in the figures on the right indicate the position of the ground state at each momentum obtained via standard exponential fits. Agreement between both methods can be seen.  The second bump in the figures on the left can be identified with the first excited state \cite{Aarts:2011sm}.
A detailed analysis of systematic uncertainties associated with MEM can be found in ref.\  \cite{Aarts:2011sm}  and a further discussion is given in appendix~\ref{sec:unc}.

\section{High temperature}
\label{sec:high}

\begin{figure}[h]
\begin{center}
\epsfig{figure=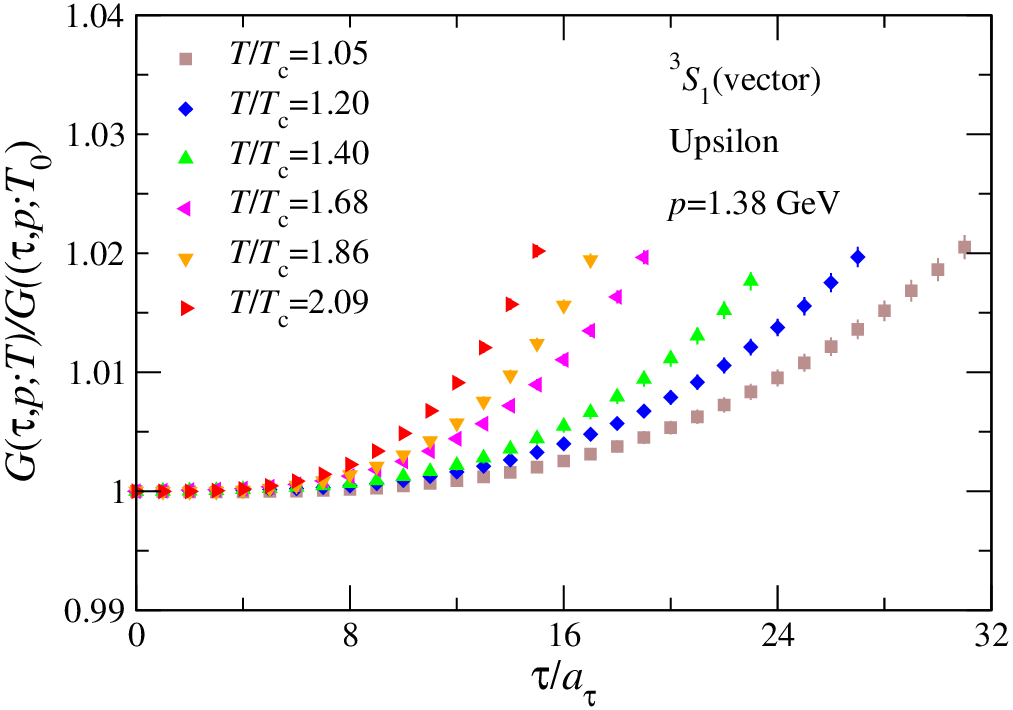,width=0.45\textwidth}
\epsfig{figure=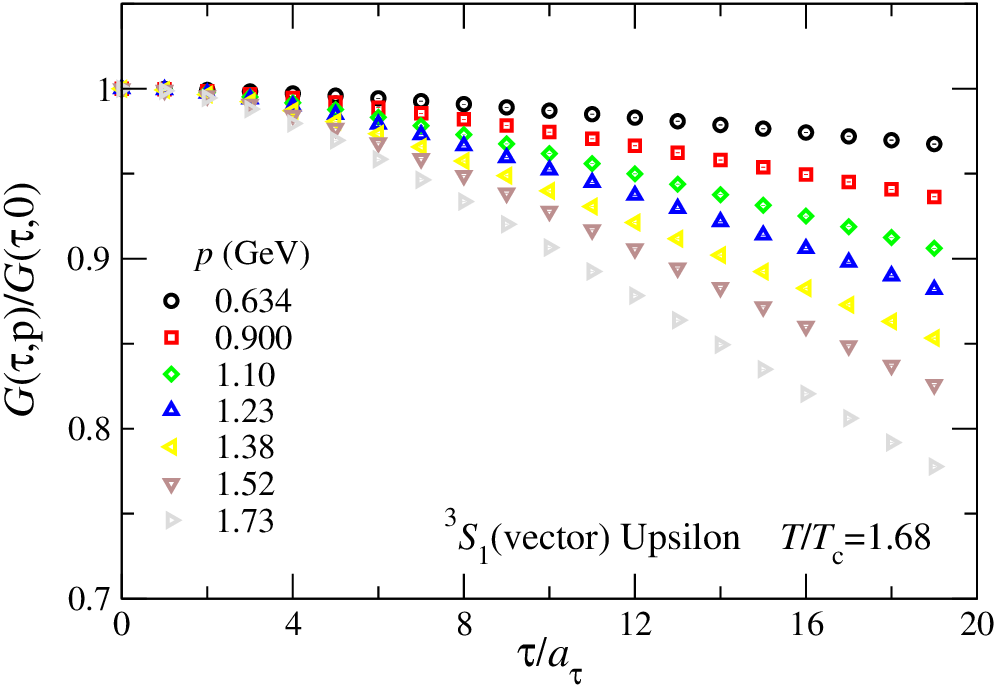,width=0.45\textwidth}
 \caption{High-temperature results in the vector ($\Upsilon$) channel. 
 Left: 
 Euclidean correlators at high temperature normalized by the correlator at the lowest temperature ($T/T_c=0.42$) at fixed momentum $p=1.38$ GeV. 
Right: 
  Euclidean correlators at nonzero momentum normalized by the correlator at zero momentum 
  at fixed temperature $T=1.68T_c$.
 }
 \label{fig:2}
 \end{center}
 \end{figure}

 We now turn to temperatures in the quark--gluon plasma phase.
  Here we find that the correlators depend on both temperature and momentum.
This is illustrated  for the vector channel in fig.\ \ref{fig:2}: on the left we present the temperature dependence via the ratio $G(\tau,\pv;T)/G(\tau,\pv;T_0)$ for fixed $p=1.38$ GeV and reference temperature $T_0=0.42T_c$, while on the right we show  the momentum dependence via the ratio $G(\tau,\pv;T)/G(\tau,\vecnul;T)$ for fixed $T=1.68T_c$.
We observe that the temperature dependence at nonzero momentum is
similar  to the one found at vanishing momentum (compare with
fig.\ 1 of ref.\  \cite{Aarts:2011sm}) and is of
the order of a few percent. Similarly the momentum dependence in the quark-gluon plasma
is similar to what we found at low temperature (compare figs.\ \ref{fig:1} and \ref{fig:2}).

\begin{figure}[t]
\begin{center}
\epsfig{figure=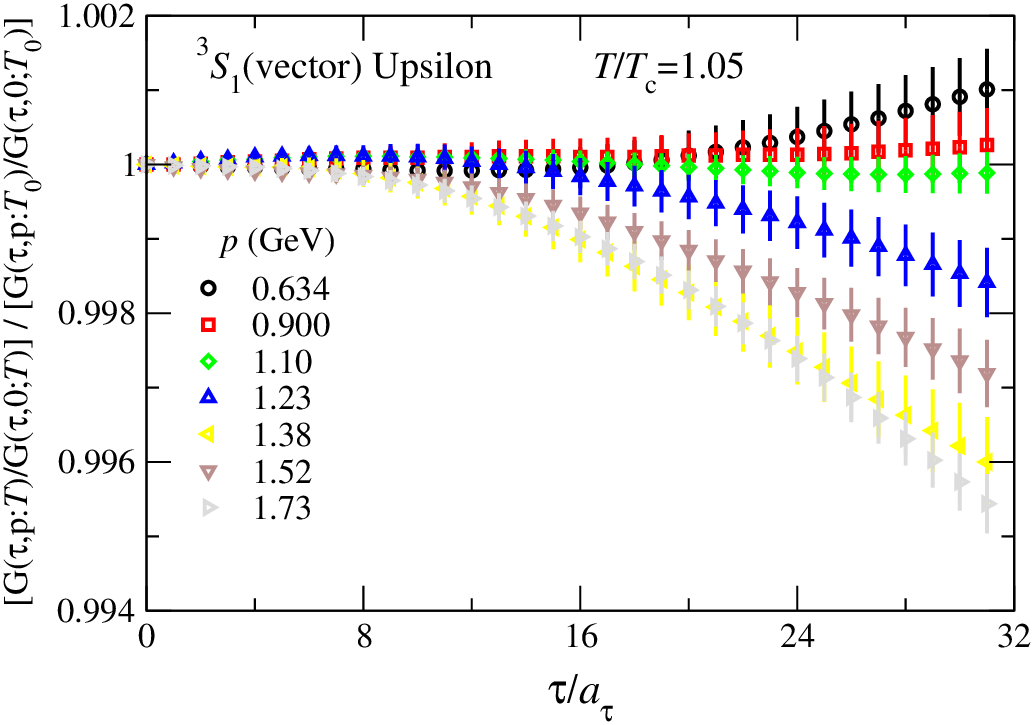,width=0.45\textwidth}
\epsfig{figure=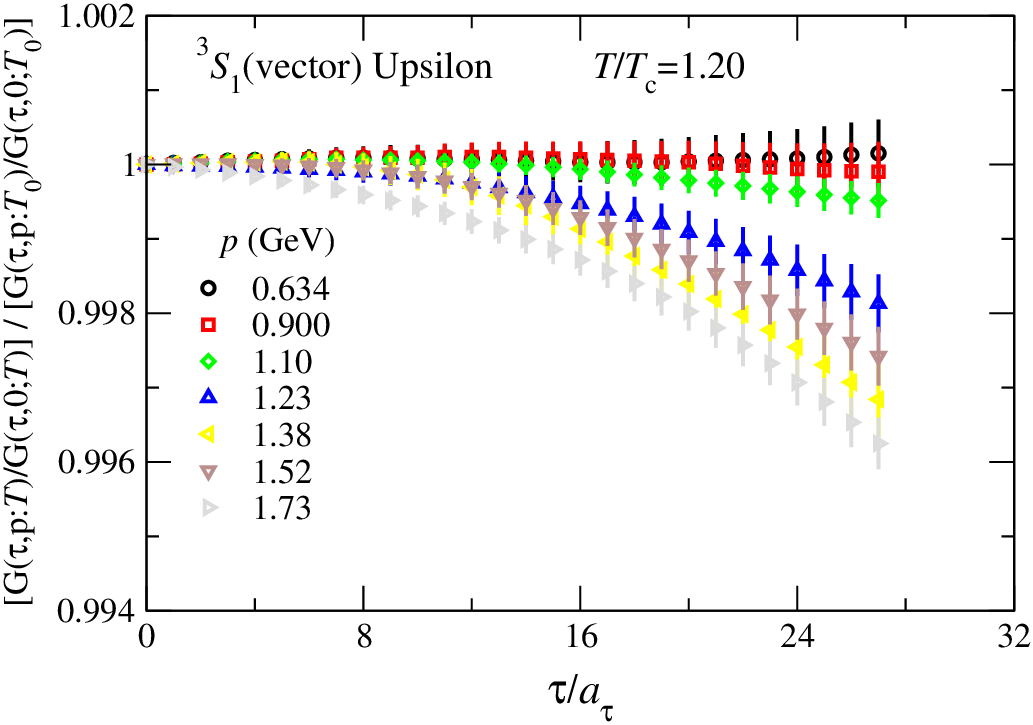,width=0.45\textwidth}\\
\epsfig{figure=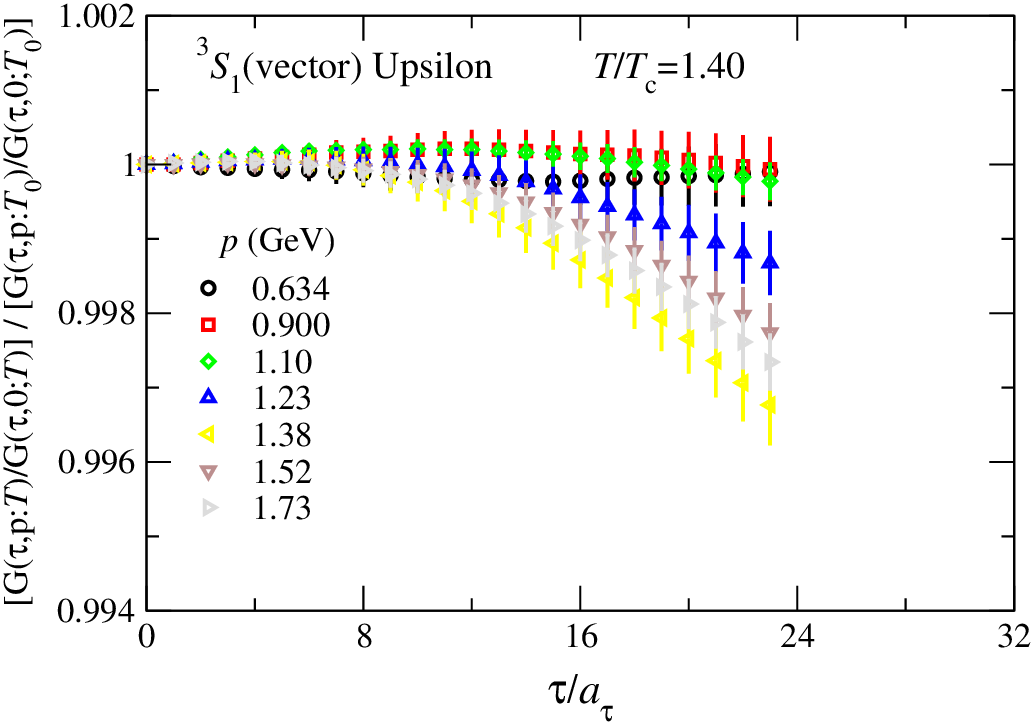,width=0.45\textwidth}
\epsfig{figure=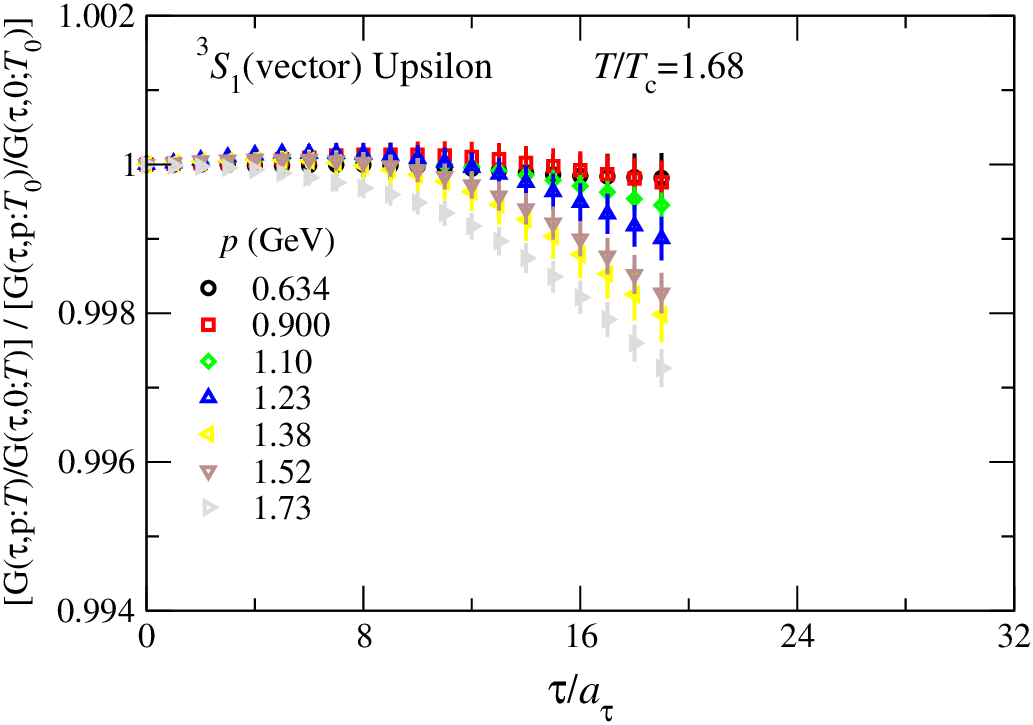,width=0.45\textwidth}\\
\epsfig{figure=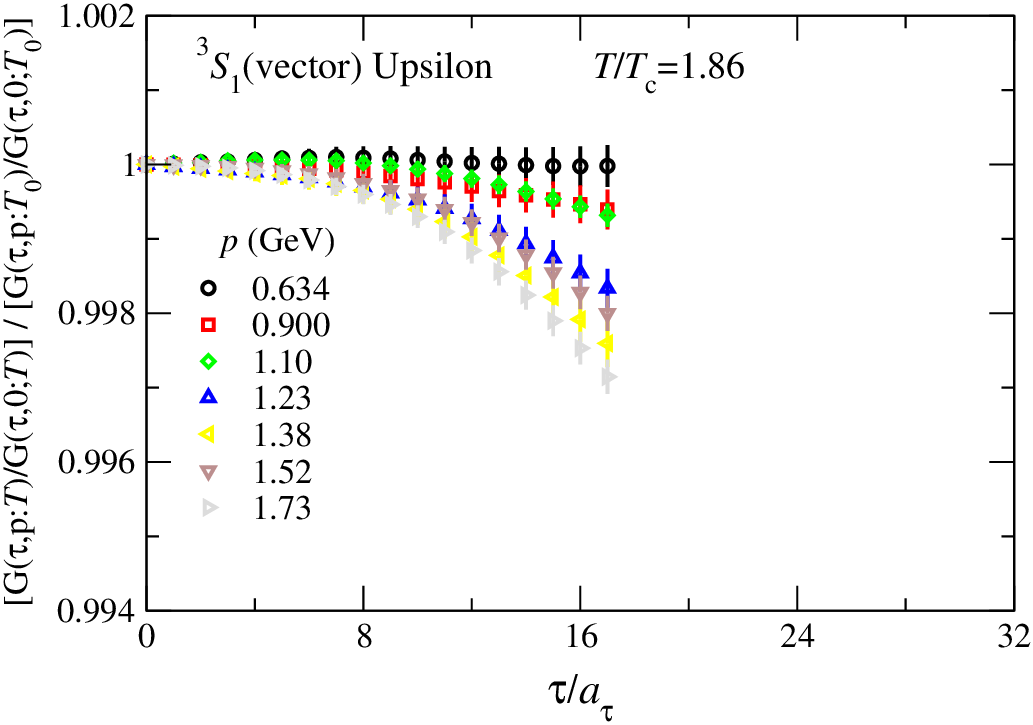,width=0.45\textwidth}
\epsfig{figure=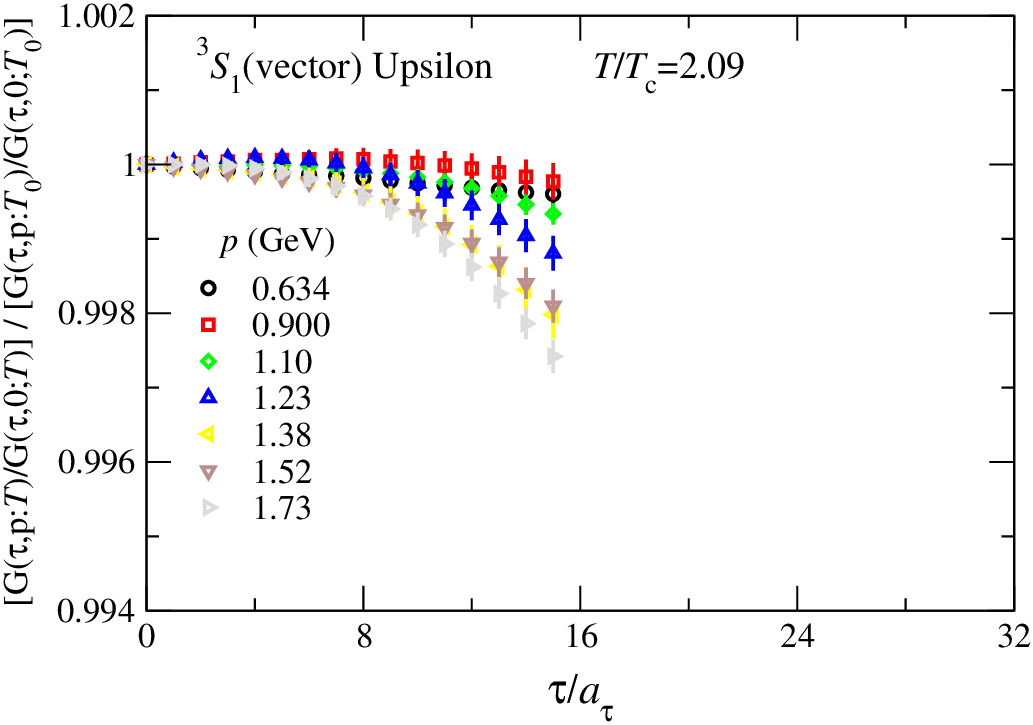,width=0.45\textwidth}
 \caption{Normalized euclidean correlators in the vector ($\Upsilon$) channel: shown is the double ratio $[G(\tau,\pv;T)/G(\tau,\vecnul;T)]/[G(\tau,\pv;T_0)/G(\tau,\vecnul;T_0)]$, where the reference temperature $T_0=0.42T_c$. Each panel shows several momenta $p$ at a fixed temperature $T$.
}
 \label{fig:Gratio-upsilon}
\end{center}
\end{figure}

In order to analyze whether the momentum dependence changes as the temperature is increased, we consider the following double ratios
\be
\frac{G(\tau,\pv;T)}{G(\tau,\vecnul;T)} \Bigg/ \frac{G(\tau,\pv;T_0)}{G(\tau,\vecnul;T_0)},
\ee
where the reference temperature is again $T_0=0.42T_c$.
The results are shown in fig.\ \ref{fig:Gratio-upsilon} in the vector channel (the pseudoscalar channel is similar). We observe that after the normalization,  the remaining momentum dependence is very mild, and always less than 0.5\% at the largest euclidean time. This apparent temperature independence of the momentum dependence provides a clear prediction which can be contrasted with potential models and EFT calculations.

\begin{figure}[t]
\begin{center}
\epsfig{figure=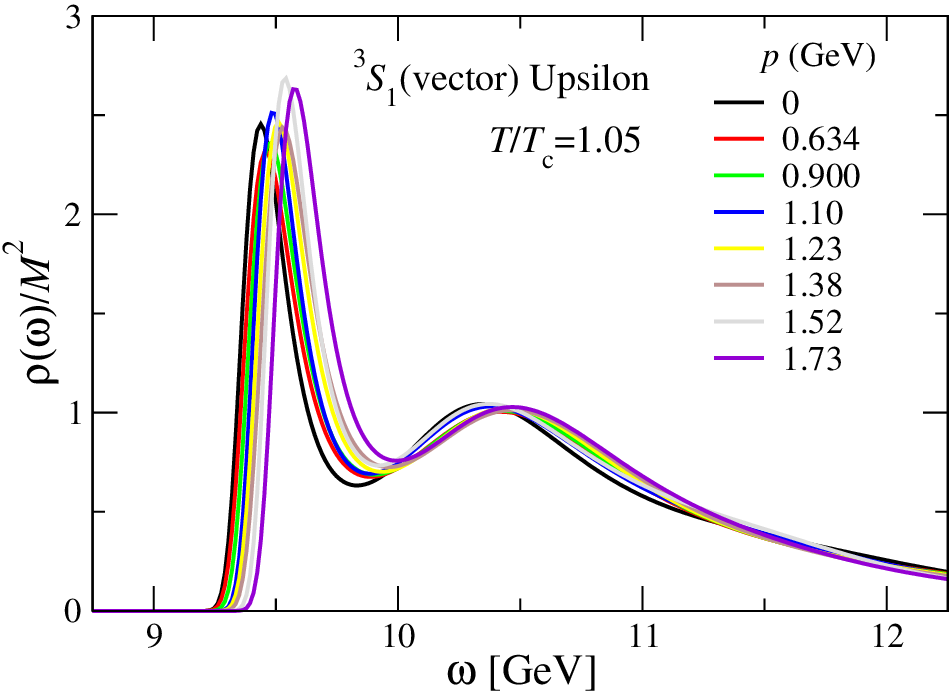,width=0.45\textwidth}
\epsfig{figure=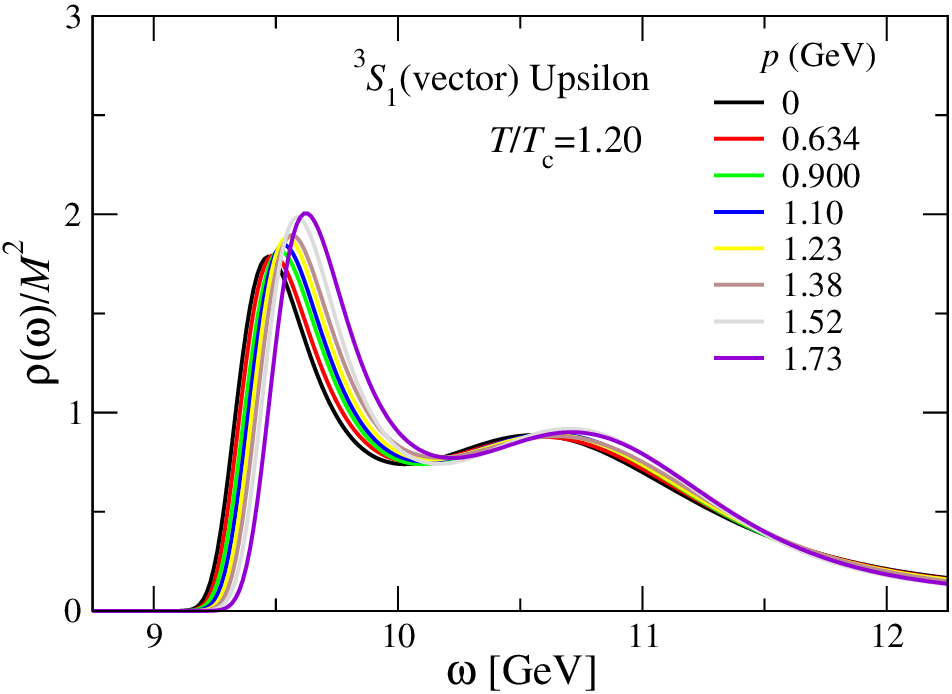,width=0.45\textwidth}\\
\epsfig{figure=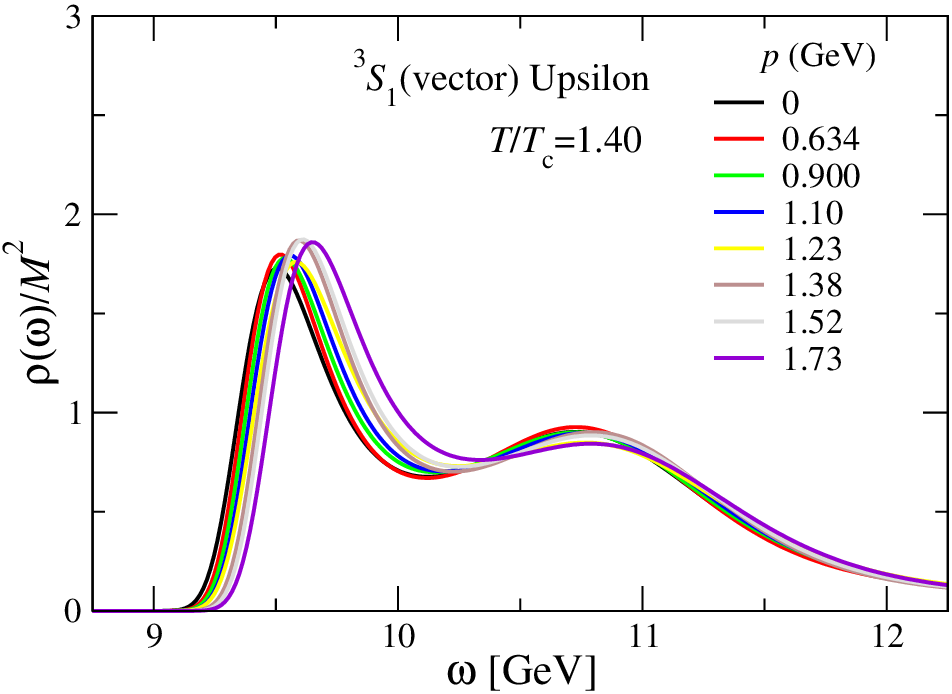,width=0.45\textwidth}
\epsfig{figure=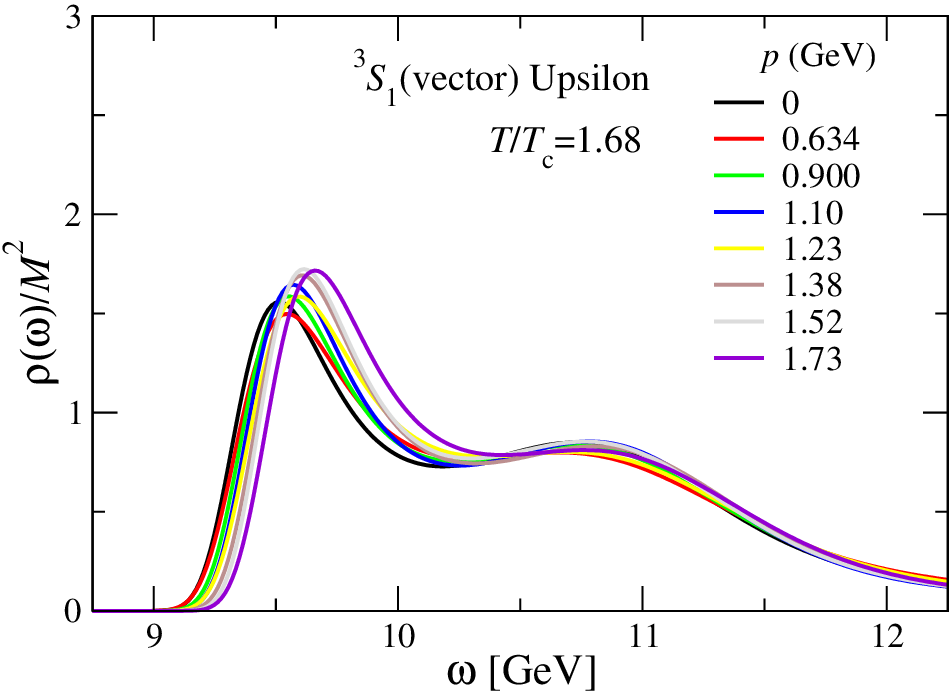,width=0.45\textwidth}\\
\epsfig{figure=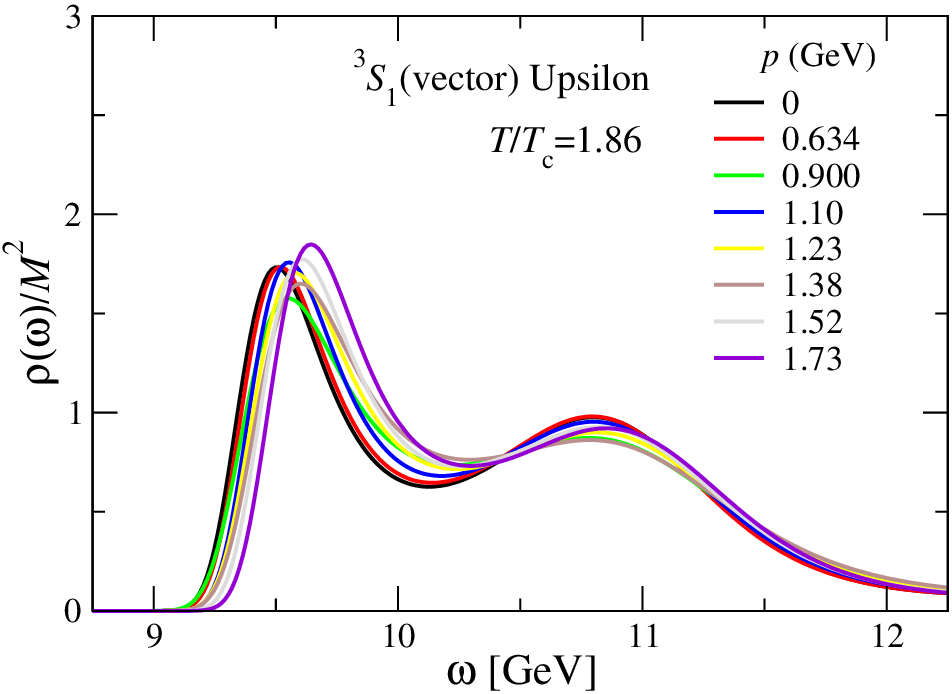,width=0.45\textwidth}
\epsfig{figure=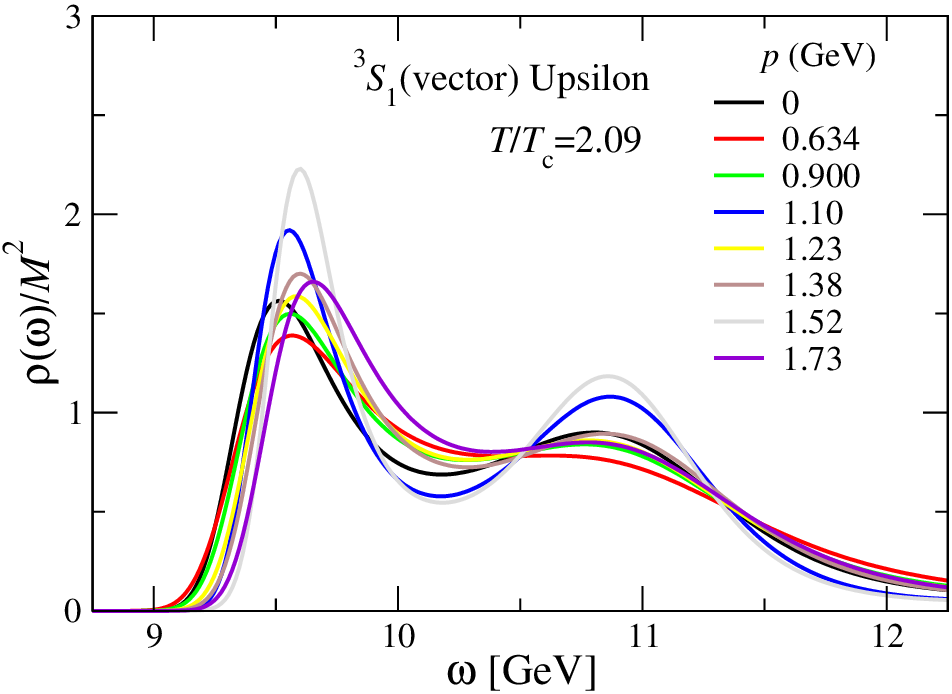,width=0.45\textwidth}
 \caption{
 Spectral functions $\rho(\om,\pv)$, normalized by the heavy quark mass, as a function of energy,
 at the six different temperatures above $T_c$  in the vector ($\Upsilon$) channel, for several momenta.
}
 \label{fig:rho-upsilon}
\end{center}
\end{figure}

\begin{figure}[t]
\begin{center}
\epsfig{figure=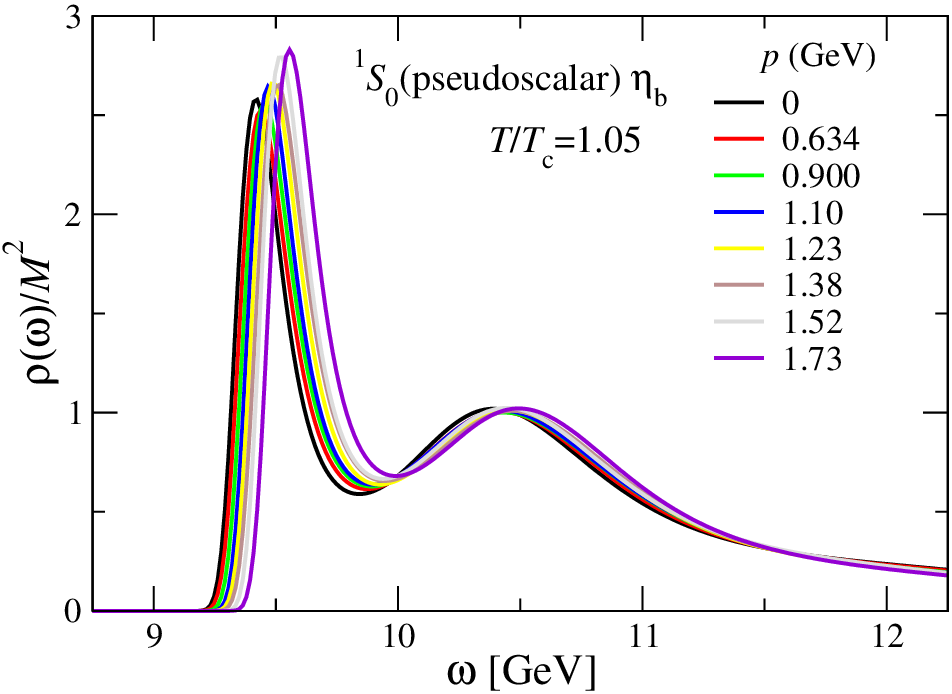,width=0.45\textwidth}
\epsfig{figure=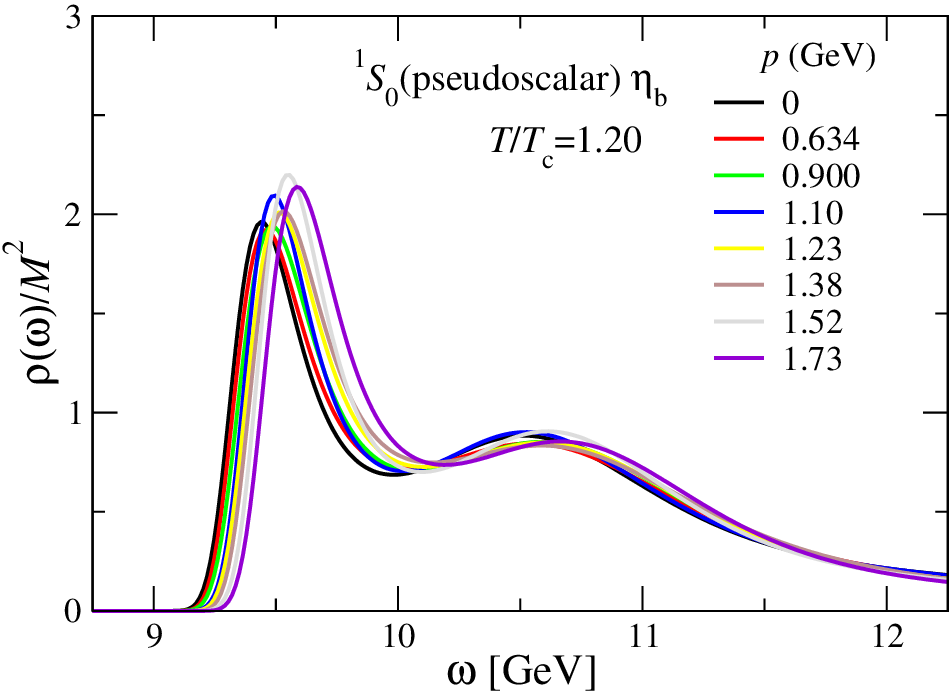,width=0.45\textwidth}\\
\epsfig{figure=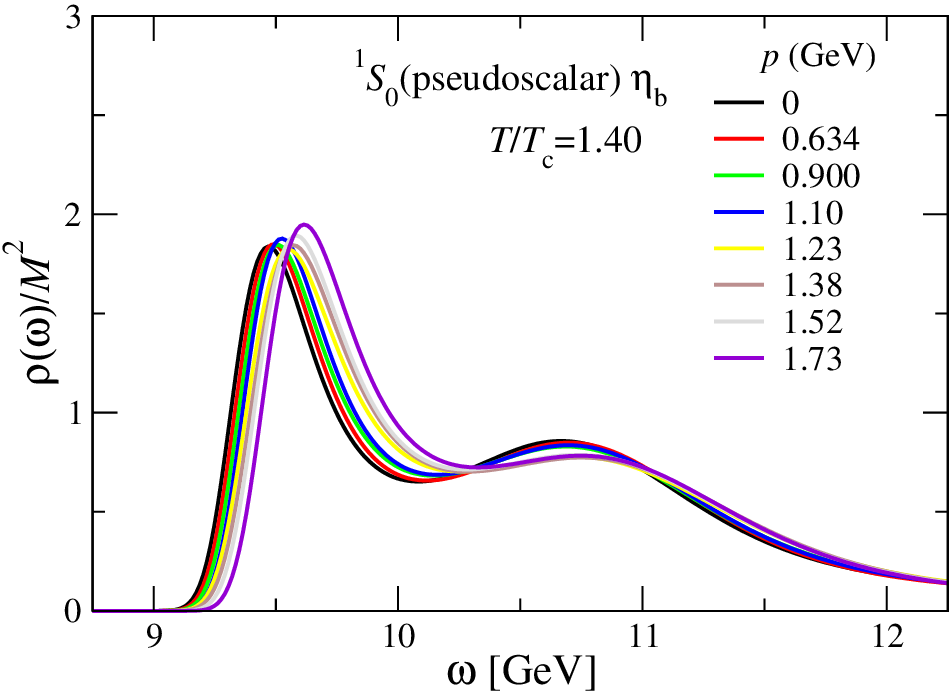,width=0.45\textwidth}
\epsfig{figure=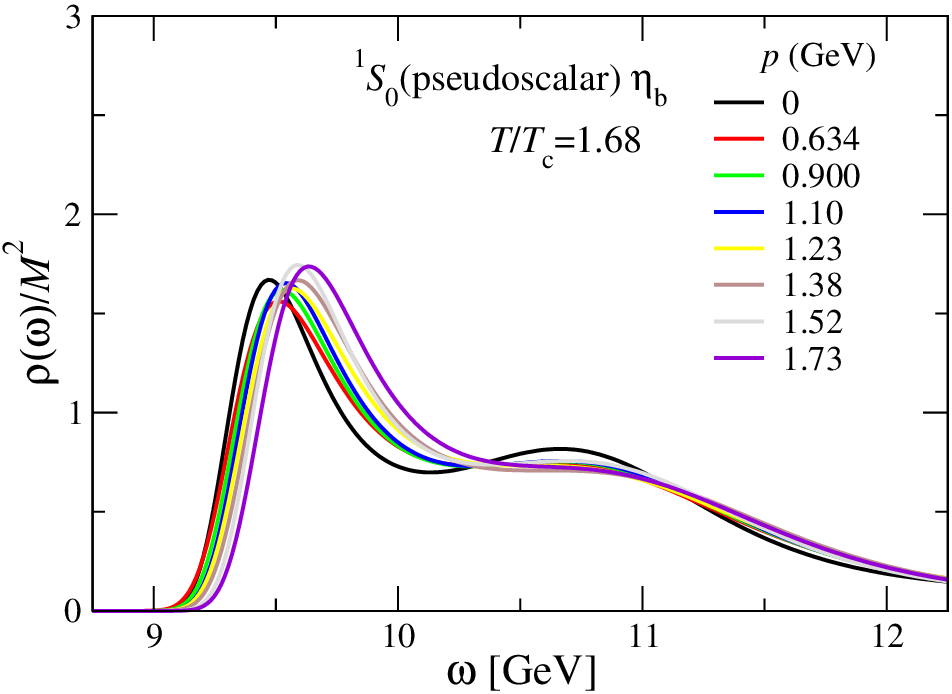,width=0.45\textwidth}\\
\epsfig{figure=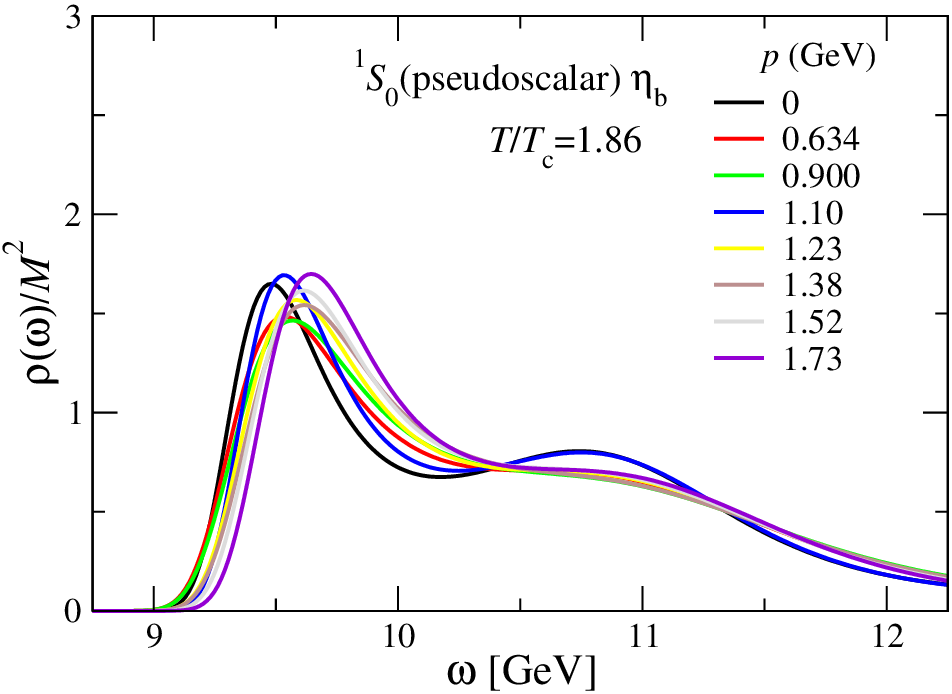,width=0.45\textwidth}
\epsfig{figure=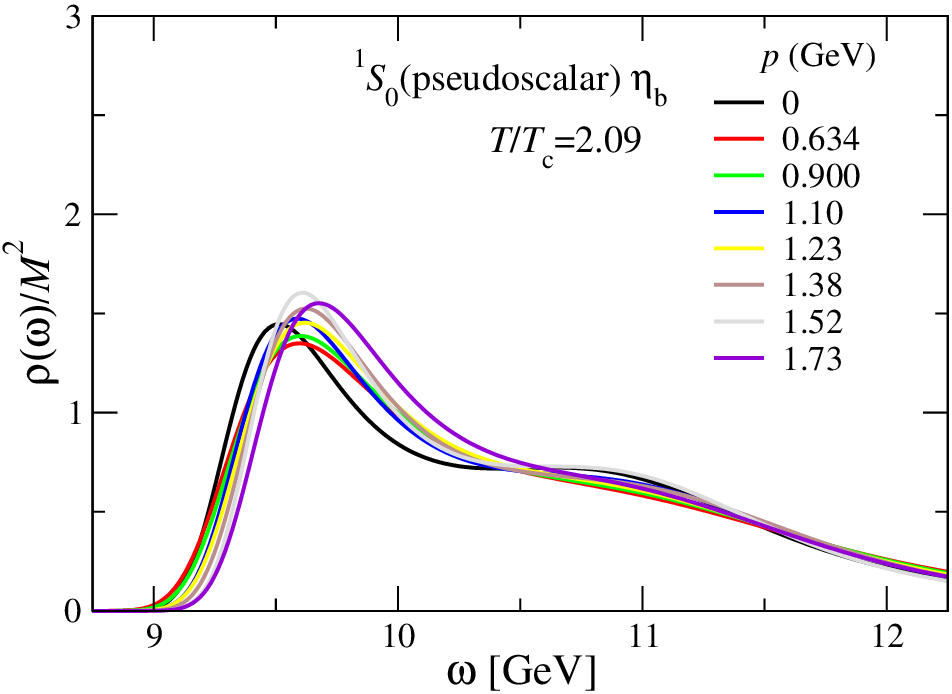,width=0.45\textwidth}
 \caption{As  fig.~\ref{fig:rho-upsilon}, for the pseudoscalar ($\eta_b$) channel.
 }
 \label{fig:rho-eta}
\end{center}
\end{figure}

The corresponding spectral functions are presented in figs.\ \ref{fig:rho-upsilon} and \ref{fig:rho-eta}.  
As expected from the results presented above and in
ref.\ \cite{Aarts:2011sm}, we find that the ground state peak moves to
the right as the momentum is increased and that it broadens and
reduces in height as the temperature is increased. As in
ref.\ \cite{Aarts:2011sm},  
  features at higher energy become suppressed as the system heats up. Note that the temporal extent becomes smaller as the temperature increases and the extraction of the spectral function more complicated. Nevertheless this may be an indication of the melting of the first excited state.
We note that the ground state peaks are very stable at low and intermediate temperatures. This is less so at the highest temperatures, presumably due the limited number of temporal lattice points available \cite{Aarts:2011sm}.

\begin{figure}[h]
\begin{center}
\epsfig{figure=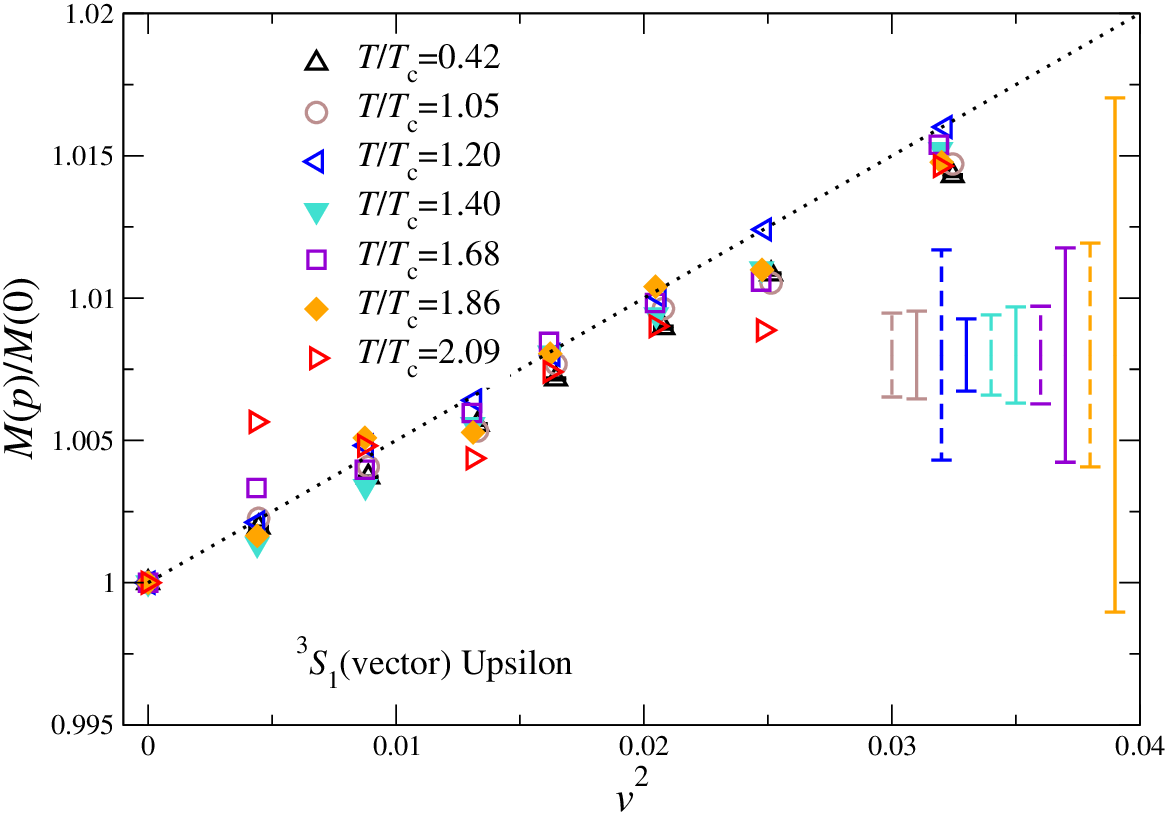,width=0.45\textwidth}
\epsfig{figure=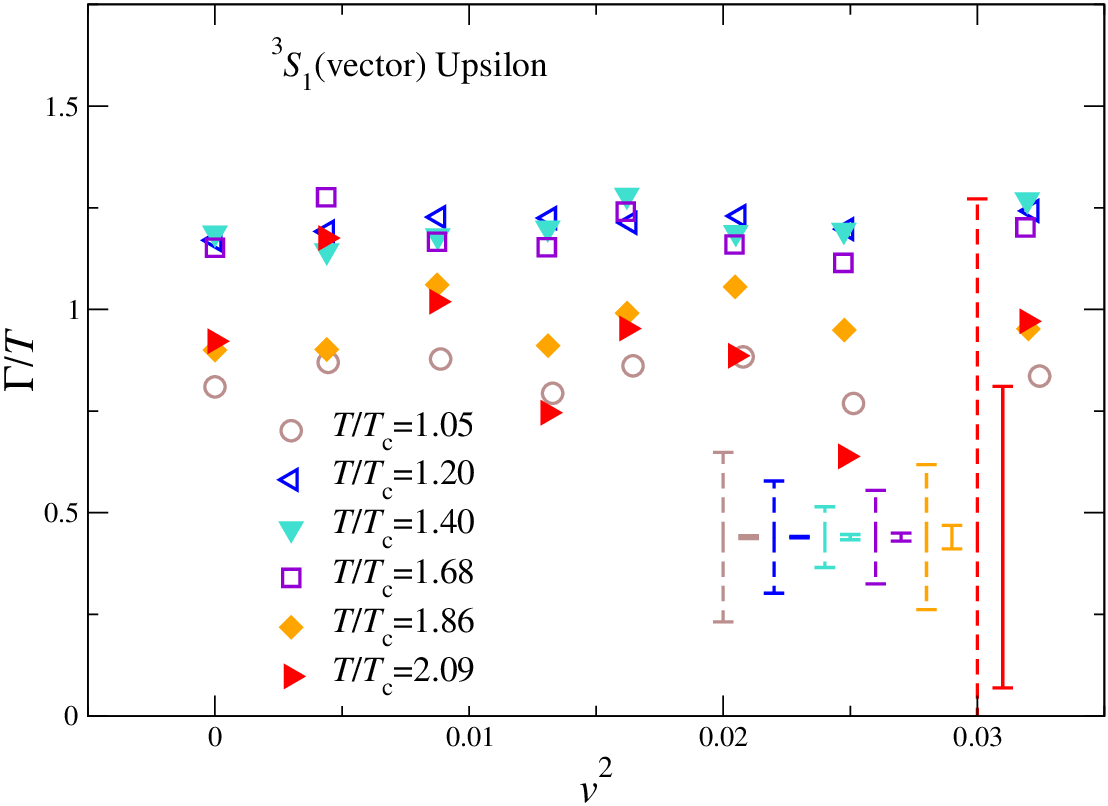,width=0.44\textwidth}
\epsfig{figure=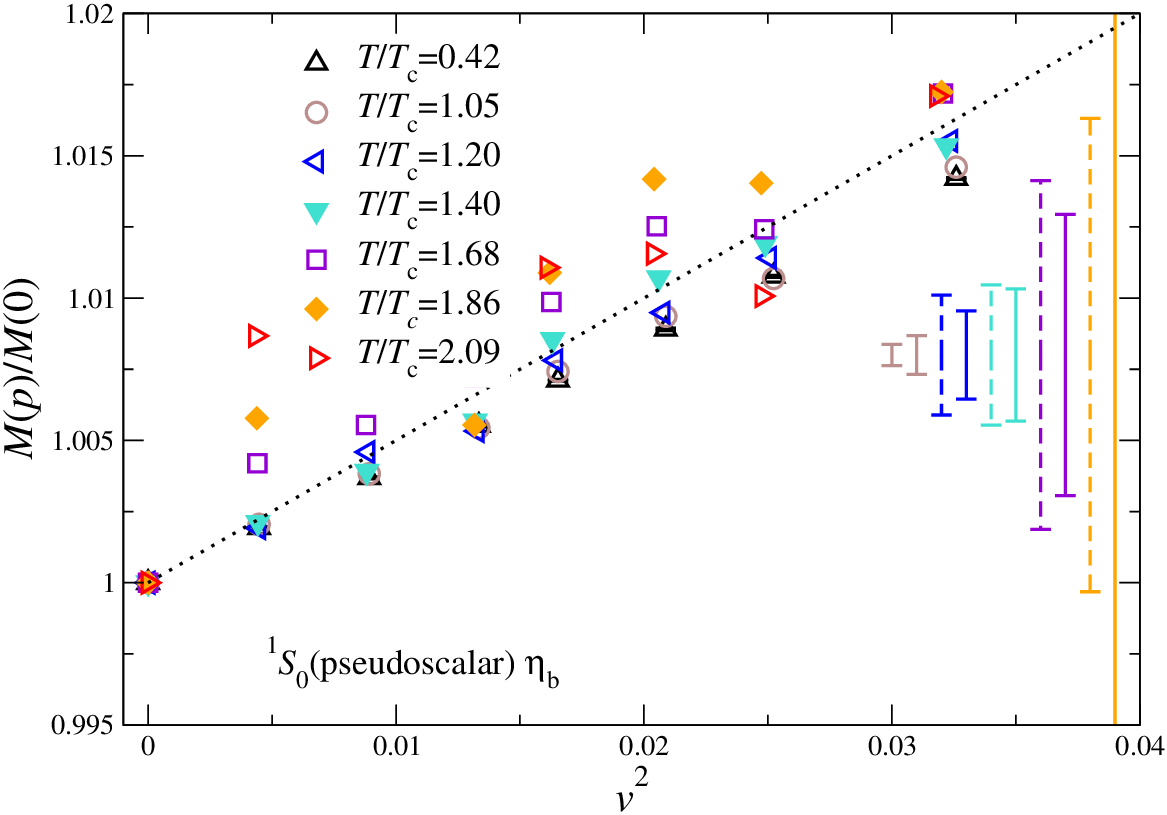,width=0.45\textwidth}
\epsfig{figure=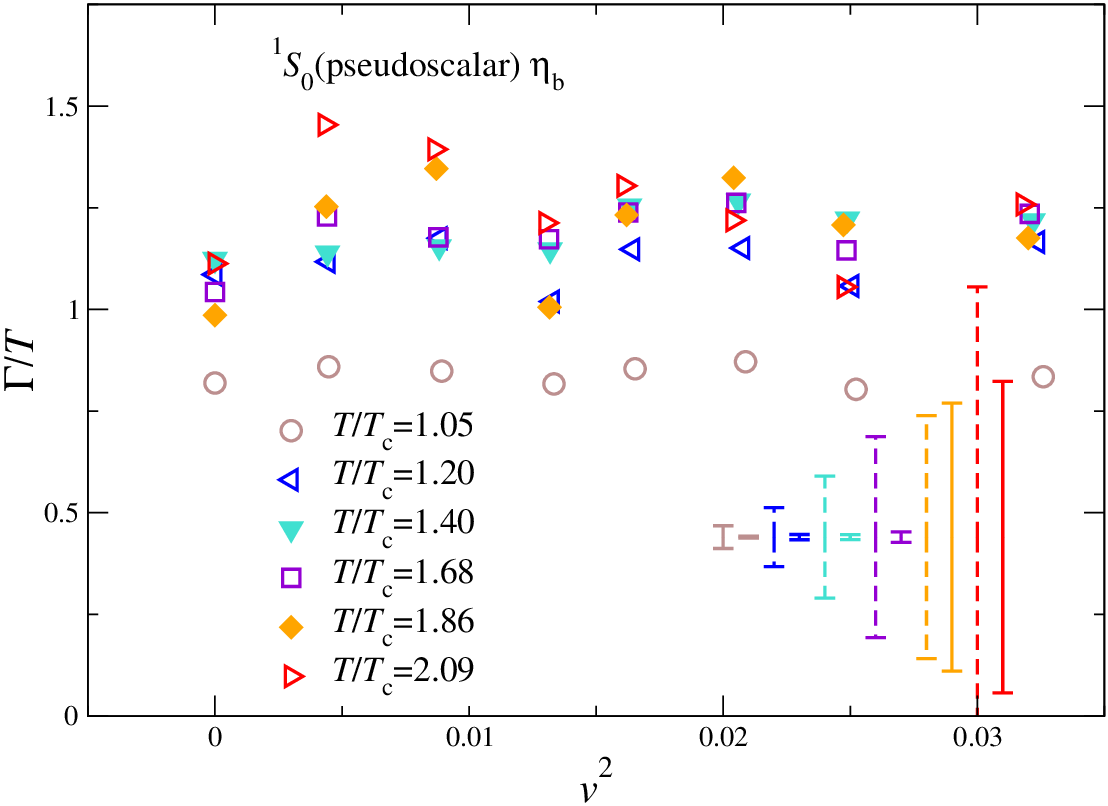,width=0.44\textwidth}
 \caption{
 Position of the ground state peak $M(\pv)/M(0)$ (left)
and  the upper limit on the width of the ground state peak, normalized
by the temperature, 
$\Gamma/T$ (right), as a function of the velocity squared ($v^2$) in the vector ($\Upsilon$) channel (above) and the pseudoscalar ($\eta_b$) channel (below), as extracted from the spectral functions. The  ground state positions below $T_c$ are obtained using a standard exponential fit. The dotted line in the left figure represents 
$M(\pv)/M(0)=1+\half v^2$.  
 The dashed error bars are representative of the systematic uncertainty at each temperature, while the solid error bars represent the statistical uncertainty; see the text for details.
 }
 \label{fig:mass-width}
 \end{center}
 \end{figure}

Following ref.\ \cite{Aarts:2011sm}, we extract the position $M(\pv)$ and width $\Gamma(\pv)$ of the ground state peak from the spectral functions. The results are shown in fig.\ \ref{fig:mass-width} in the vector and  pseudoscalar channels, as a function of $v^2$, where $v=|\pv|/M(\vecnul, T)$, with $M(\vecnul,T)$ the peak position at zero momentum and temperature $T$.
The temperature dependence of $M(\vecnul,T)$ has been discussed and compared with EFT calculations in ref.\ \cite{Aarts:2011sm}.
 The peak position below $T_c$ has been extracted using standard exponential fits, but agrees with the one obtained from the spectral functions, see figs.\ 
 \ref{fig:rho-upsilon-80} and \ref{fig:rho-eta-80}.
 Note that the width is normalized by the temperature. 
 Uncertainties enter in various ways: we refer to the uncertainty due to the finite number of time slices that are included in the MEM analysis as the systematic uncertainty and due to the finite number of configurations used in the MEM computation as the statistical uncertainty \cite{Aarts:2011sm}. Both depend on the temperature and velocity;  to avoid cluttering
we show for each temperature the largest systematic and statistical error, with the left one (dashed) representing systematic and the right one (full) representing statistical uncertainties.  For the mass, error bars are not shown for the highest temperature as they exceed the graph size.
Errors are under control at the lower temperatures, but uncertainties become large at the highest temperatures. The position of the peak is easier to determine than its width. We also note that at the lower temperatures the uncertainty in the width is dominated by the systematic error, whereas for the mass the two contributions are of similar magnitude.
In appendix \ref{sec:unc} we show that there is no default model dependence for the ground state peak.

We observe that the peak position increases linearly with $v^2$, as expected. Assuming the lowest-order, nonrelativistic expression $M(\pv)=M(0) + \pv^2/2M(0)$, one finds
\be
\frac{M(\pv)}{M(0)} = 1+ \frac{\pv^2}{2M^2(0)} = 1+\half v^2,
\ee
which is indicated with the dotted lines in the left figures.
The slope of the data points at the lowest temperature is slightly less than $1/2$, which can be improved by a careful tuning of the heavy quark mass.
Systematic uncertainties are too large to find a trend in the velocity dependence of the width; 
we note that the width is independent of the velocity within errors.

The dependence on the velocity  can be compared with EFT predictions. In ref.\ \cite{Escobedo:2011ie} a study of the velocity dependence was carried out  in the context of QED, working in the rest frame of the bound state (i.e.\ the heat bath is moving). In order to compare with our setup, we consider the case in which the temperature is low enough for bound states to be present and that the velocities are nonrelativistic. In that case, one finds \cite{Escobedo:2011ie}, in the rest frame of the bound state and at leading order in the EFT expansion, 
\be
\frac{\Gamma_v}{\Gamma_0}  = \frac{\sqrt{1-v^2}}{2v}\log\left(\frac{1+v}{1-v}\right),
\ee
where $\Gamma_0$ is the width at  rest.
Interpreting the width as an inverse lifetime, one can express this
result  in the rest frame of the heat bath by dividing with the Lorentz factor $\gamma = 1/\sqrt{1-v^2}$. An expansion for nonrelativistic velocities then yields
\be
\frac{\Gamma_v}{\Gamma_0} =   1 -\frac{2v^2}{3} +  {\cal O}\left(v^4\right).
\ee
If we take this result and apply it to our study of bottomonium,  we find that 
the effect of the nonzero velocity shows up as a correction at the percent level (recall that $v^2\lesssim 0.04$), which is beyond our level of precision but consistent with the observed $v$ independence within errors.
Similarly, additional thermal effects in the dispersion relation are currently beyond our level of precision.

\section{Summary and outlook}
\label{sec:sum}

In this paper we extended our analysis of bottomonium in the quark-gluon plasma to nonzero  
momentum, using lattice QCD simulations with two flavours of light quarks and nonrelativistic 
dynamics for the bottom quark. We analyzed both the euclidean correlators as well as the associated 
spectral functions constructed  using the maximum entropy method. While we observed both 
temperature and momentum dependence in the correlators and spectral functions, we found that the 
momentum dependence is effectively temperature independent. This is seen directly in the 
correlators as well as in the position and width of the ground state and can be compared with existing 
and future EFT predictions. 

As an outlook we plan to improve the tuning of the heavy quark mass,
with the aim to enhance the prospects of detecting thermal deviations
from the standard dispersion relation. We are  planning to carry out
this analysis  in a quark-gluon plasma with $N_f=2+1$ rather
than 2 flavours, with both a smaller spatial lattice spacing and a larger spatial extent,  which will also allow us to reach higher velocities.

We hope that this work provides further encouragement to study
quarkonia at nonzero momentum using EFT and potential model approaches
and can contribute to the interpretation of the experimental results for bottomonium in heavy ion collisions at the LHC.

\acknowledgments

GA and JIS thank Joan Soto for discussion.
 GA and JIS thank the Department of Energy's Institute for Nuclear Theory at the University of Washington for its hospitality and the Department of Energy for partial support during stages of this work.
 GA, CRA and MPL thank the Galileo Galilei Institute for Theoretical Physics for the hospitality and the INFN for partial support during the completion of this work.
 We acknowledge the support and
infrastructure provided by the Trinity Centre for High Performance
Computing and the IITAC project funded by the HEA under the Program
for Research in Third Level Institutes (PRTLI) co-funded by the Irish
Government and the European Union.  The work of CA and GA is carried
as part of the UKQCD collaboration and the DiRAC Facility jointly
funded by STFC, the Large Facilities Capital Fund of BIS and Swansea
University.  GA and CA are supported by STFC.  SK is grateful to STFC
for a Visiting Researcher Grant and supported by the National Research
Foundation of Korea grant funded by the Korea government (MEST) No.\
2012R1A1A2A04668255.  SR is supported by the Research Executive Agency (REA)
of the European Union under Grant Agreement number PITN-GA-2009-238353
(ITN STRONGnet) and the Science Foundation Ireland, grant no.\
11-RFP.1-PHY-3201.  DKS is supported in part by US Department of
Energy contract DE-AC02-06CH11357.  JIS  has been supported by Science
Foundation Ireland grant 08-RFP-PHY1462  and 11-RFP.1-PHY-3193-STTF11.

\appendix

\section{Free NRQCD spectral functions at nonzero momentum}
\label{sec:free}
\setcounter{equation}{0}

In this appendix we generalize the results of appendix A in Ref.\ \cite{Aarts:2011sm} for free nonrelativistic quarks to nonzero momentum, following closely the analysis in Ref.\  \cite{Burnier:2007qm}. Consider heavy quarks in (continuum) NRQCD with energy $E_\kv=\kv^2/2M$. The 
correlators for the $S$ and $P$ waves at nonzero spatial momentum $\pv$ are then~\cite{Burnier:2007qm}
 \begin{align}
\label{eq:GS}
 G_{S}(\tau,\pv)  = &\, 2N_c \int \frac{d^3k}{(2\pi)^3}\, e^{-(E_\kv+E_{\pv+\kv})\tau} 
 = e^{-\tau p^2/4M} G_{S}(\tau,\vecnul), \\ 
\label{eq:GP}
 G_{P}(\tau,\pv)  = &\, 2N_c \int \frac{d^3k}{(2\pi)^3}\, (\kv+\pv/2)^2 e^{-(E_\kv+E_{\pv+\kv})\tau}
 = e^{-\tau p^2/4M} G_{P}(\tau,\vecnul),
 \end{align}
where we shifted the momentum $\kv\to \kv-\pv/2$ to isolate the external momentum dependence.
Recall that
 \be
 G_{S}(\tau, \vecnul)  = \frac{N_c}{4\pi^{3/2}}\left(\frac{M}{\tau}\right)^{3/2}, 
\quad\quad\quad
 G_{P}(\tau, \vecnul)  = \frac{3N_c}{8\pi^{3/2}} \left(\frac{M}{\tau}\right)^{5/2}.
\ee
Interestingly, for free quarks nonzero momentum changes a powerlaw decay into an exponential decay. 

In terms of spectral densities, using
\be
G(\tau, \pv) = \int_{\om_{\rm min}}^{\om_{\rm max}} \frac{d\om}{2\pi}\, e^{-\om\tau}\rho(\om, \pv),
\ee
this results in a simple shift of the threshold \cite{Burnier:2007qm}
\begin{align}
\label{eq:rhoS}
 \rho_{S}(\om, \pv)  = &\,  4\pi N_c \int \frac{d^3k}{(2\pi)^3}\, \delta\left(\om'-2E_\kv\right)
 = \frac{N_c}{\pi} M^{3/2} \left(\om'\right)^{1/2}\Theta(\om'),
  \\ 
\label{eq:rhoP}
 \rho_{P}(\om, \pv)  = &\,  4\pi N_c \int \frac{d^3k}{(2\pi)^3}\, \kv^2 \delta\left(\om'-2E_\kv \right) 
 = \frac{N_c}{\pi} M^{5/2} \left(\om'\right)^{3/2}\Theta(\om').
 \end{align}
 where $\om'=\om-p^2/4M$.
 
Similarly, the only modification to the free lattice spectral functions  shown in Ref.\ \cite{Aarts:2011sm} due to the nonzero momentum is the shift of $\omega$ according to $\om'=\om-p^2/4M$.

\section{More details on the maximum entropy method}
\label{sec:unc}
\setcounter{equation}{0}

In this appendix we provide some further details on the maximum entropy method regarding default model dependence and error estimates, extending the discussion of ref.~\cite{Aarts:2011sm}.

\begin{figure}[h]
\centerline{
  \includegraphics[width= 0.45\textwidth]{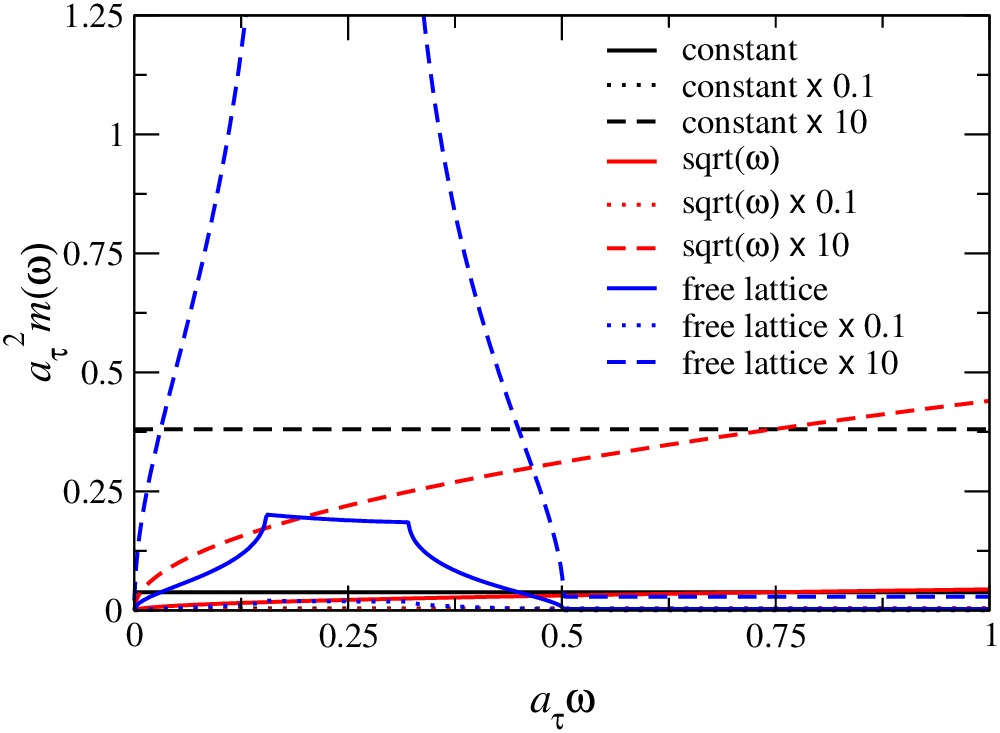}
  }
 \caption{Default models $m(\om)$ used in the analysis.
  }
\label{fig:default1}
\end{figure}

 In  ref.\ \cite{Aarts:2011sm} we demonstrated default model independence using a large set of default models. However, we did not include default models based on free lattice spectral functions, so this is done here. 
We employ default models $m(\om)$ specified by $m(\om)/m_0= 1$, $\sqrt{\om}$, $m_{\rm lattice}(\om)$, where the latter is the free lattice spectral function discussed in appendix \ref{sec:free} and in ref.\ \cite{Aarts:2011sm}. The parameter $m_0$ is specified by $m_0/m_0^*=0.1,1,10$, with $m_0^*$ determined by minimizing $\chi^2$ which appears in the MEM approach \cite{Aarts:2011sm}. Since MEM requires the default model to be nonzero in the $\omega$ interval of interest,  we added a small constant to the lattice default model, whenever it is zero, i.e.\ when $a_\tau\om\gtrsim 0.5$.  These default models are shown in fig.\ \ref{fig:default1}.

\begin{figure}[h]
\centerline{
  \includegraphics[width= 0.45\textwidth]{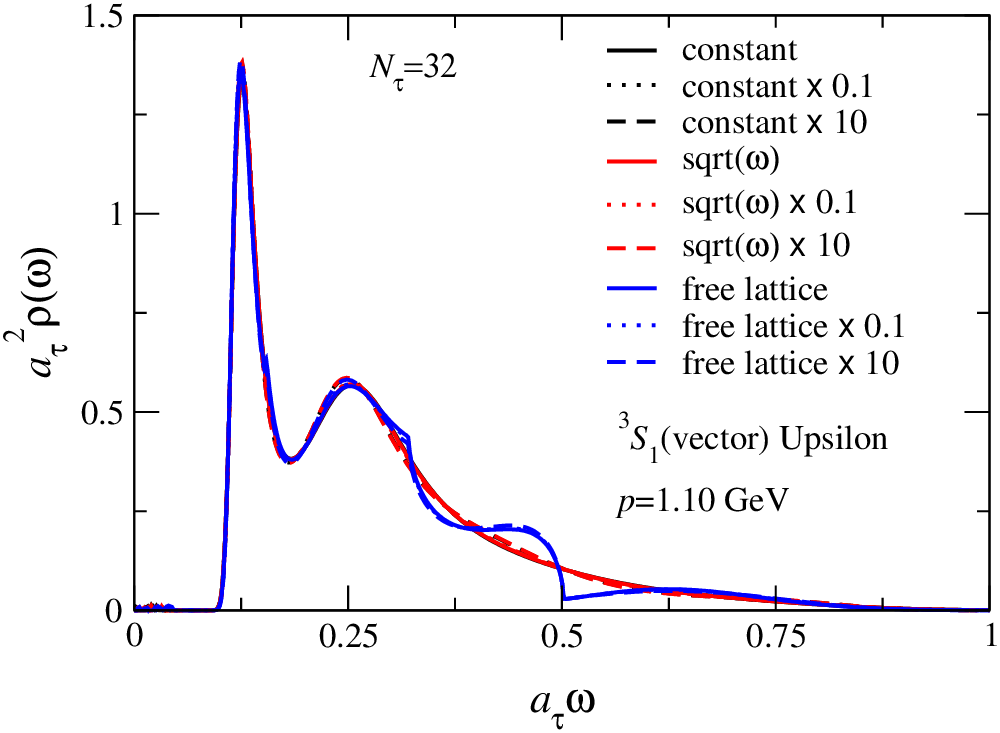}
    \includegraphics[width= 0.45\textwidth]{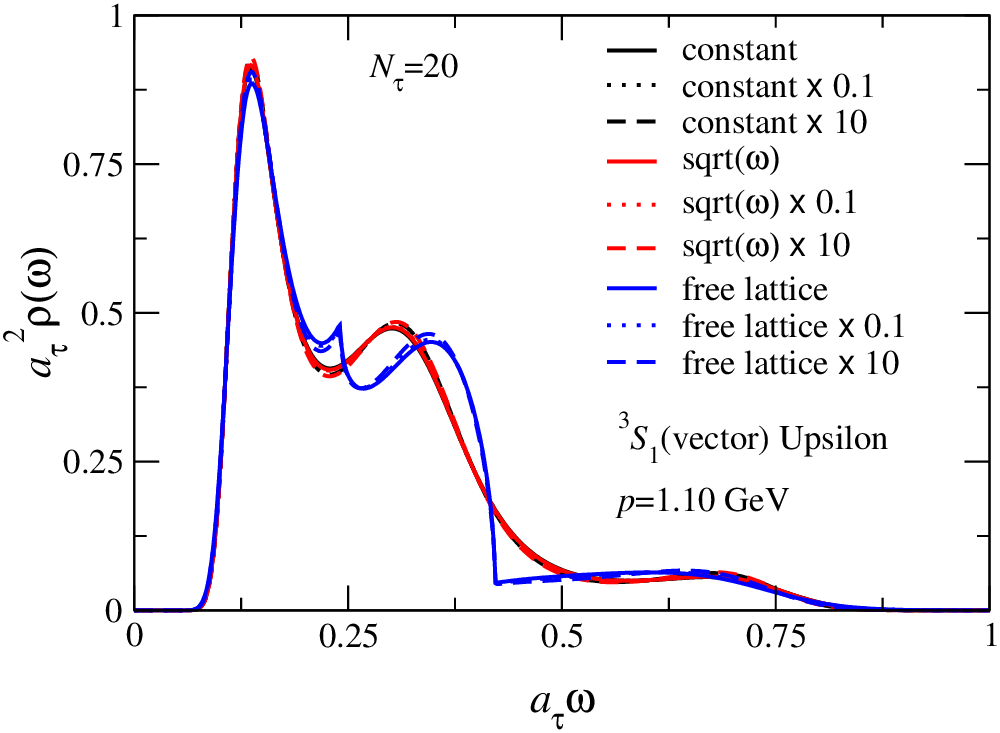}
  }
 \caption{Default model dependence at $N_\tau=32$ (left) and 20 (right) at $p=1.10$ GeV.
   }
\label{fig:default2}
\end{figure}

The resulting default model dependence is shown in fig.\ \ref{fig:default2} for $N_\tau=32$ and 20 at $p=1.10$ GeV in the vector channel. We observe that the ground state peak is independent of the default model. The results from the smooth default models (i.e.\ $m(\om)\sim$ constant, $\sqrt{\om}$) agree. 
The free lattice default model has cusps at larger energies, related to the edges of the Brillouin zone. This nonanalyticity is still visible in the constructed spectral functions, although they roughly follow the trend of the other ones. At higher temperature (smaller $N_\tau$), there is more sensitivity to the lattice default model, but not to the other ones.

\begin{figure}[h]
\begin{center}
\epsfig{figure=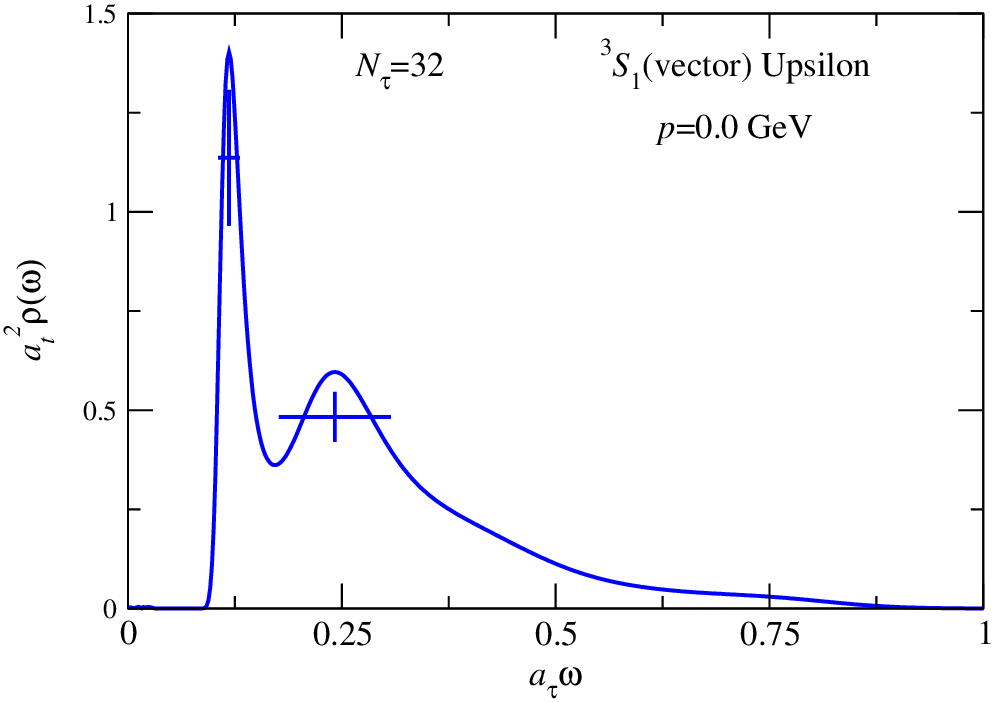,width=0.45\textwidth}
\epsfig{figure=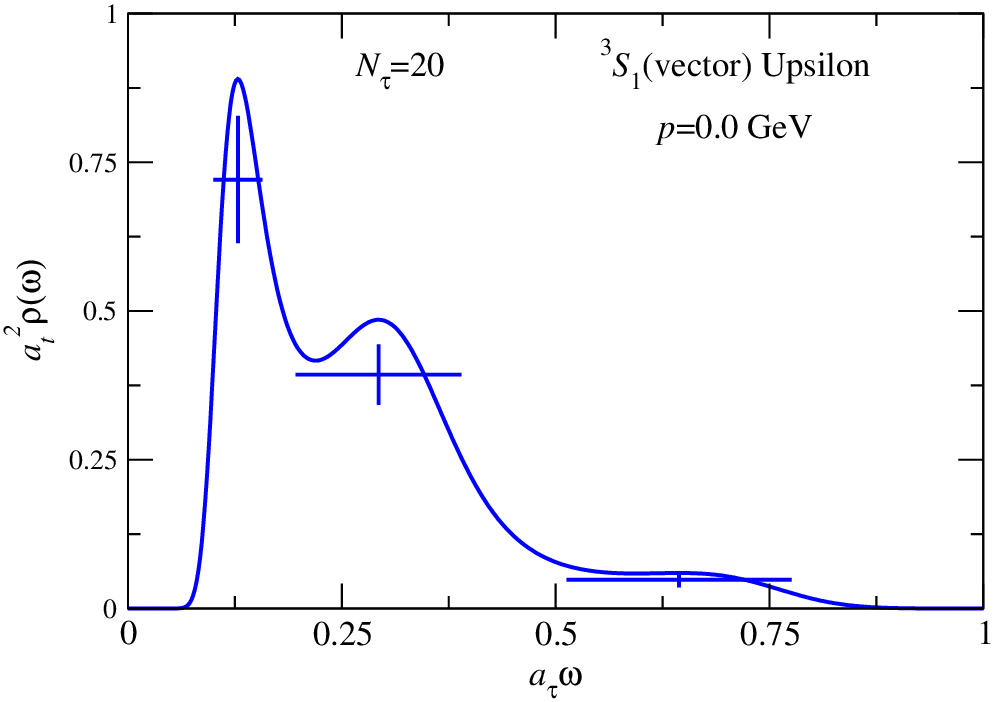,width=0.45\textwidth}
\epsfig{figure=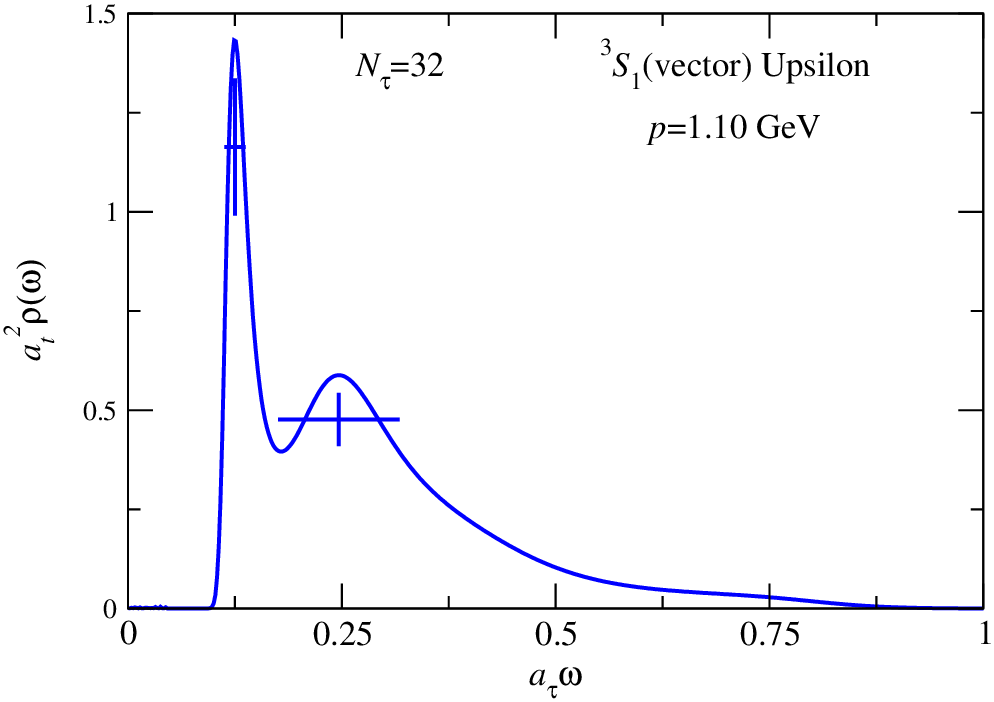,width=0.45\textwidth}
\epsfig{figure=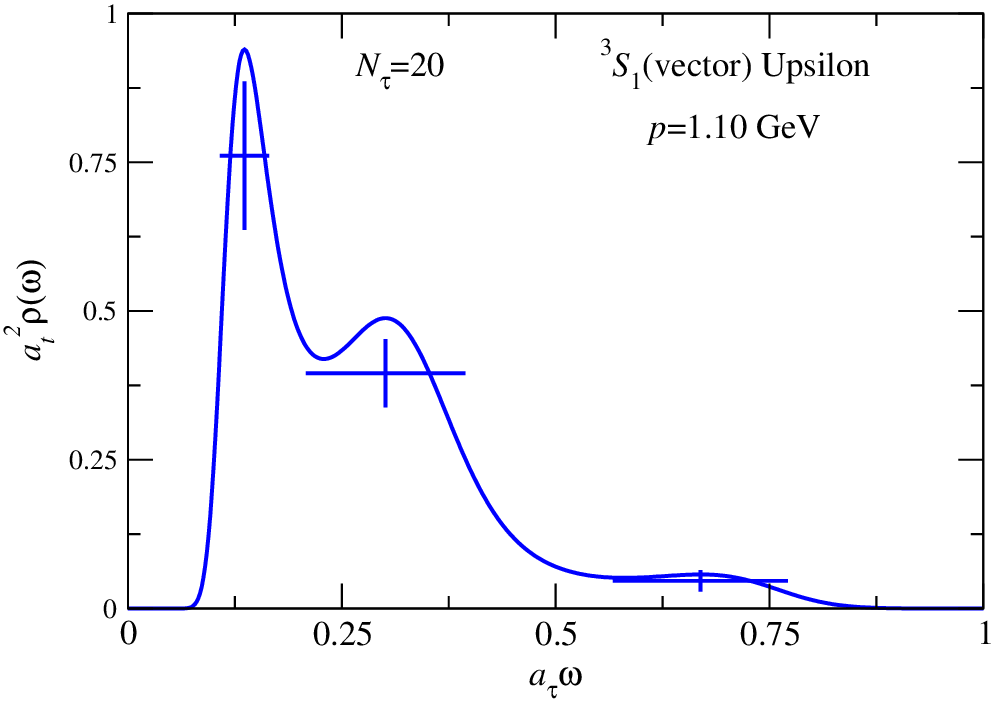,width=0.45\textwidth}
 \caption{
Significance of spectral features at $N_\tau=32$ (left) and 20 (right), at $p=0$ (top) and 1.10 (bottom) GeV. The vertical error bars represent the error in the average height of the spectral feature in the region indicated by the horizontal error bar. The vertical position of the error bar indicates the average height in that region. The horizontal bar does not indicate the uncertainty in the peak position.
  }
 \label{fig:err}
\end{center}
\end{figure}

Uncertainties in spectral features can also be assessed by estimating the statistical significance of the peaks. Following the standard MEM analysis, see e.g.\ ref.~\cite{Aarts:2007pk}, we determine the uncertainty in the average peak height within a given $\omega$ interval. Some representative results are shown in fig.\ \ref{fig:err}, at zero and nonzero momentum. Here the vertical error bar represents the error in the average height of the spectral feature as determined by the MEM procedure, where the average was performed over the interval indicated with the horizontal error bar. The vertical coordinate of the horizontal error bar equals the average height itself. The horizontal bar is centred at $M$ and extends over the interval $[M - \Gamma/2$, $M + \Gamma/2]$, where $M$ and $\Gamma$ are the position and width of the feature.
We emphasize that  the width of the horizontal bar does not correspond to the error in the peak's position, but to the interval that is considered.
We conclude that the ground state peak is a well-defined feature at all temperatures. The second peak is seen to become less significant as the temperature is increased, in agreement with our previous conclusions~\cite{Aarts:2011sm}.

\end{document}